# Gate-Tunable Reversible Rashba-Edelstein Effect in a Few-Layer Graphene/2H- TaS$_2$ Heterostructure at Room Temperature

Lijun Li[1], Jin Zhang[2], Gyuho Myeong[1], Wongil Shin[1], Hongsik Lim[1], Boram Kim[1], Seungho Kim[1], Taehyeok Jin[1], Stuart Cavill[3], Beom Seo Kim[3], Changyoung Kim[4], Johannes Lischner[2], Aires Ferreira[3]*, Sungjae Cho[1]*

We report the observation of current-induced spin polarization, the Rashba-Edelstein effect (REE), and its Onsager reciprocal phenomenon, the spin galvanic effect (SGE), in a few-layer graphene/2H-TaS$_2$ heterostructure at room temperature. Spin-sensitive electrical measurements unveil full spin-polarization reversal by an applied gate voltage. The observed gate-tunable charge-to-spin conversion is explained by the ideal work function mismatch between 2H-TaS$_2$ and graphene, which allows strong interface-induced Bychkov-Rashba interaction with a spin-gap reaching 70 meV, while keeping the Dirac nature of the spectrum intact across electron and hole sectors. The reversible electrical generation and control of the nonequilibrium spin polarization vector, not previously observed in a nonmagnetic material, are elegant manifestations of emergent 2D Dirac fermions with robust spin-helical structure. Our experimental findings, supported by first-principles relativistic electronic structure and transport calculations, demonstrate a route to design low-power spin-logic circuits from layered materials.

[1]Department of Physics, Korean Advanced Institute of Science and Technology (KAIST), Daejeon, Korea
[2]Departments of Materials and Physics, Imperial College London, SW7 2AZ, United Kingdom
[3]Department of Physics, University of York, YO105DD, United Kingdom
[4]Department of Physics, Seoul National University, Seoul, Korea
*email: sungjae.cho@kaist.ac.kr
*email: aires.ferreira@york.ac.uk

The enhancement of spin-orbit effects at interfaces with broken inversion symmetry holds great promise for the development of spin-logic technologies that can offer high-speed operation with reduced energy consumption[1-3].

Graphene is considered a promising two-dimensional (2D) material for next-generation spintronics[4], owing to its micrometer spin diffusion length[5,6] and electrically tunable electronic structure[7,8]. Recently, the family of transition metal dichalcogenides (TMDs)—layered crystals with large spin-orbit coupling (SOC)[9]—has enlarged the breadth of accessible SOC phenomena in graphene-based heterostructures, by enabling all-optical spin injection[10,11] and strong proximity-induced SOC up to 17 meV[14-16], more than 100 times greater than graphene's intrinsic spin-orbit gap[17]. Microscopically, TMDs induce symmetry-distinct SOCs in graphene. The enhancement of $\mathbf{z} \rightarrow -\mathbf{z}$ mirror-symmetric SOC[18,19] is predicted to induce spin Hall effect (SHE)[20-24], whereby an applied charge current $\mathbf{J}$ is converted into a transverse spin current $\mathbf{J}_s^z = (\hbar/2e)\, \alpha_{\mathrm{SH}}\, \hat{\mathbf{z}} \times \mathbf{J}$, where $\alpha_{\mathrm{SH}}$ is the spin Hall angle and $\hat{\mathbf{z}}$ is the unit vector normal to the basal plane. Simultaneously, the interfacial breaking of inversion symmetry in graphene/TMD heterostructures induces Bychkov-Rashba interaction[25,26]. The emergence of $\mathbf{z} \rightarrow -\mathbf{z}$ asymmetric SOC (hereafter, referred to as Rashba SOC) is predicted to entangle spin and SU(2)-sublattice-pseudospin degrees of freedom, endowing the 2D Dirac states with a Fermi-energy-dependent helical spin texture in momentum space[27,28], a graphene counterpart of spin-momentum-locked surface states in topological insulator thin films.

Interfacial states with spin-helical structure allow efficient charge-spin interconversion via the Rashba-Edelstein effect (REE)[29] and its Onsager reciprocal phenomenon, the spin galvanic effect (SGE)[30]. In the REE, an applied charge current magnetizes the 2D conduction electrons, generating a nonequilibrium spin density

$$\delta \mathbf{S} = (\hbar/2e)\, \beta_{EE}\, \hat{\mathbf{z}} \times \mathbf{J}, \quad (1)$$

where $\beta_{EE}$ is the microscopic figure of merit for charge-to-spin conversion[28]. The latter is directly related to the experimentally accessible charge-to-spin conversion efficiency $\gamma_{REE}$, defined as the ratio of the spin- and charge-current densities in the linear response regime. Observations of REE or SGE have been reported for quantum wells[31], metallic bilayers[32,33], oxide heterostructures[34,35], and topological insulator/metal heterostructures[36], and also recently in graphene/semiconducting TMD heterostructures[37,38].

Theoretical analysis of coupled charge-spin transport in graphene/TMD heterostructures has shown that spin-helical 2D Dirac fermions enable robust room-temperature REE that can be controlled by a gate voltage. In contrast to charge-to-spin conversion induced by surface states of topological insulators[36], the sign of the REE in graphene/TMD heterostructures depends on the charge carrier polarity[28]. The figure of merit in ideal (particle-hole symmetric) conditions satisfies $\beta_{EE}(n) = -\beta_{EE}(-n)$, where $n = n(V_g)$ is the charge carrier density induced by a gate voltage. Recent experiments indicate that REE in heterostructures of graphene and semiconducting TMDs is accompanied by SHE[37,38], in agreement with the general symmetry relations dictating that the interplay of Rashba SOC and proximity spin-valley coupling generates robust SHE[23], while only mildly affecting the REE efficiency[28]. However, the expected gate modulation of the REE signal has only been detected at low temperatures for electron-type carriers ($n > 0$)[37] indicating that the spin-helical structure in $\mathbf{k}$-space is not fully established. Moreover, the strength of interfacial SOC in graphene/semiconducting TMD heterostructures is on the same order of the typical quasiparticle broadening in clean graphene (1-10 meV)[37,38], which limits the accessible REE conversion rates[28].

The discovery of a graphene/metallic TMD heterostructure enabling strong interfacial SOC with effective gate control over both electron and hole-type carriers is thus necessary to realize the full potential of graphene for the development of energy-efficient spintronic devices.

Here, we used the metallic layered compound 2H-TaS$_2$ (group V dichalcogenide) to induce Rashba SOC in graphene. The work function of 2H-TaS$_2$ ($W_{TaS_2} \approx 5.6$ eV[39]) is very close to the predicted critical value[40] where repulsive chemical interactions between the metallic TMD and graphene precisely balance the driving force for charge transfer arising from the work function difference of the two subsystems. This enables a strong interface-induced SOC, while minimizing the Fermi level shift with respect to the unperturbed Dirac ($K$) points, thus providing ideal conditions to explore the interplay of spin and pseudospin degrees of freedom, which is manifest in the carrier density dependence of main transport quantities. Figures 1a and b show, respectively, a schematic illustration and an optical image of the device. A quasi-one-dimensional spin diffusion channel of an exfoliated five-layer graphene (5LG) flake is used to propagate spin currents between the SOC-active heterojunction (5LG/2H-TaS$_2$), where charge-spin conversion takes place, and the ferromagnetic contact FM3, which plays the role of a spin probe (spin injector) in the REE (SGE) measurement scheme. Due to the additional screening of impurity potentials by the multiple layers, few-layer graphene typically displays enhanced spin coherence in comparison to monolayer graphene[41]. A section (3.6 × 6.5 μm$^2$) sof 5LG is covered with a 15nm-thick 2H-TaS$_2$ strip to create the 5LG/2H-TaS$_2$ heterojunction, where the interfacial SOC is induced in the carbon layer. As shown below, the minimal charge transfer between 5LG and 2H-TaS$_2$ enables gate control over spin transport phenomena across both electron and hole bands. Details on the device fabrication are given in Supplementary Information (SI).

# Figure 1| Characterization of the 5LG/2H-TaS₂ heterostructure device.

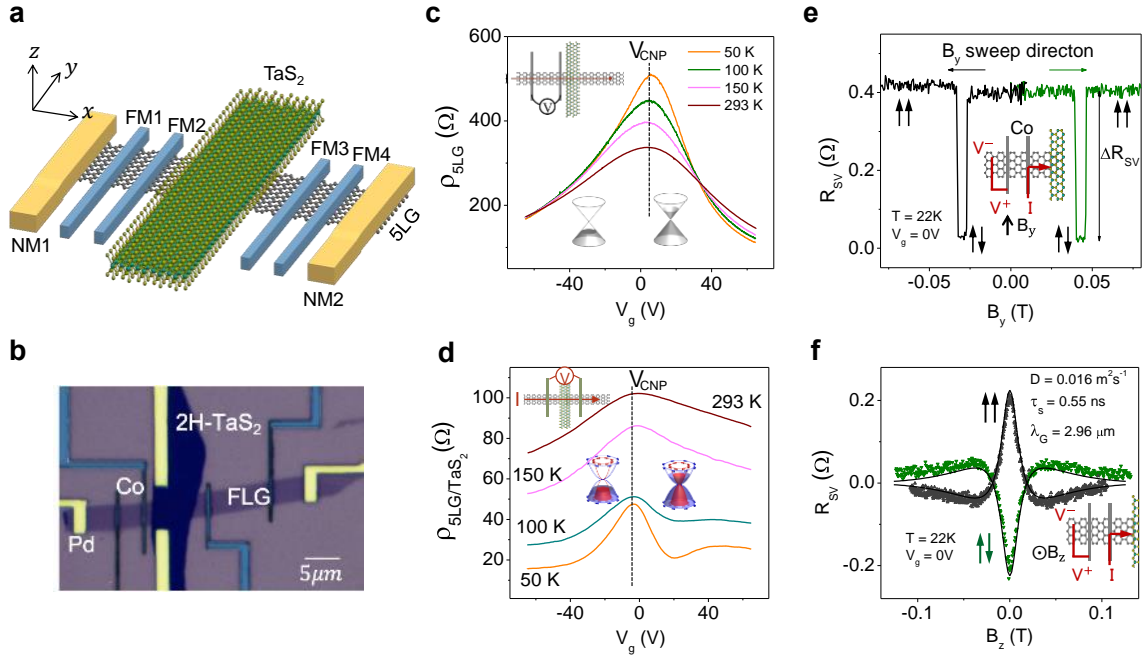

**(a)** Schematic illustration of the van der Waals REE device. The device contains nonmagnetic ohmic electrodes (Cr/Pd, 3/45 nm, labeled NM1 and NM2) and ferromagnetic contacts (Co/Pd, 50/5 nm, labeled FM1-FM4). The electrodes FM1-2 (located at distances $L_1 = 5$ μm and $L_2 = 1$ μm from the heterojunction) are used for spin transport characterization of the 5LG flake not covered by 2H-TaS2. **(b)** False colored optical image of the device. **(c)** 2D resistivity $\rho_{5LG}$ of 5LG at selected temperatures. The temperature dependence shows the expected charge transport transition from a localized ($d\rho/dT < 0$) regime near the CNP to a metallic regime ($d\rho/dT > 0$) at high $V_g$. **(d)** 2D resistivity $\rho_{5LG/TaS_2}$ of the heterojunction at selected temperatures. The 5LG/TaS2 temperature dependence shows metallic behavior over the entire gate voltage range. **(e)** Nonlocal spin-valve resistance $R_{SV}$ measured in magnetic field $\boldsymbol{B} = B_y\, \hat{\boldsymbol{y}}$ at $V_g = 0$. The green (black) data correspond to a positive (negative) field sweep. Vertical arrows in (e) and (f) indicate parallel/antiparallel configurations of the ferromagnetic electrodes' magnetization. **(f)** Nonlocal Hanle spin precession measurement with magnetic field $\boldsymbol{B} = B_z\, \hat{\boldsymbol{z}}$ at $V_g = 0$. Points are experimental data. The diffusion constant $D$ and the in-plane spin lifetime $\tau_s$ of the 5LG are extracted from a fit to the theoretical Hanle spin precession curve (solid lines).

First, we perform control measurements to characterize the individual device components of our device (5LG spin diffusion channel, 5LG/2H-TaS2 and 2H-TaS2). Figure 1c shows the zero-field square resistance of 5LG, $\rho_{5LG}$, as a function of gate voltage. $\rho_{5LG}$ attains a maximum at the charge neutrality point (CNP) $V_g \approx 7$ V, and decreases monotonically with increasing charge density for both carrier polarities. The temperature dependence shows a from an insulating regime near the CNP to a metallic regime at high $V_g$, consistent with the semi-metallic nature of Bernal-stacked multilayer graphene.[42,43] The 2D resistivity of the 5LG/2H-TaS2 heterojunction, $\rho_{5LG/TaS_2}$, exhibits fine gate tunability akin to $\rho_{5LG}$ (Fig. 1d). This is in contrast to the two-probe resistance of $2H\text{-}TaS_2$, which remains constant over the entire $V_g$ range as expected for a metallic TMD (SI). The CNP shifts from $V_g \approx 7$ V in 5LG to $V_g \approx -3$ V at the heterojunction (see Figs. 1c-d). This shows that at $V_g = 0$, the Fermi level in 5LG/2H-TaS2 remains close to the conical Dirac (*K*) point of graphene. This feature is attributed to the ideal work function mismatch between 2H-TaS2 and 5LG, consistent with the general charge transfer model for graphene by Giovannetti et al.[40], and confirmed for 5LG/2H-TaS2 by means of dedicated density functional theory calculations (see below). Crucially, $\rho_{5LG/TaS_2}$ shows metallic behavior from 50 K up to room temperature, while allowing effective gate tunability up to room temperature. These results indicate that the interfacial states of 5LG, albeit strongly perturbed by 2H-TaS2, dominate the charge carrier transport in the heterojunction region.

To characterize the spin transport in 5LG, lateral spin-valve measurements were carried out. Figure 1e shows the measured nonlocal spin-valve resistance $R_{SV}$ as function of the magnetic field $B_y$ applied along common easy axis of the ferromagnetic probes $T = 22$ K and $V_g = 0$ V. The abrupt changes in $R_{SV}$ correspond to the magnetization switching of the ferromagnetic electrodes from antiparallel to parallel configurations. To assess the spin transport fidelity of the 5LG channel

**Figure 2| Measurement schemes for REE and SGE and Onsager reciprocity**

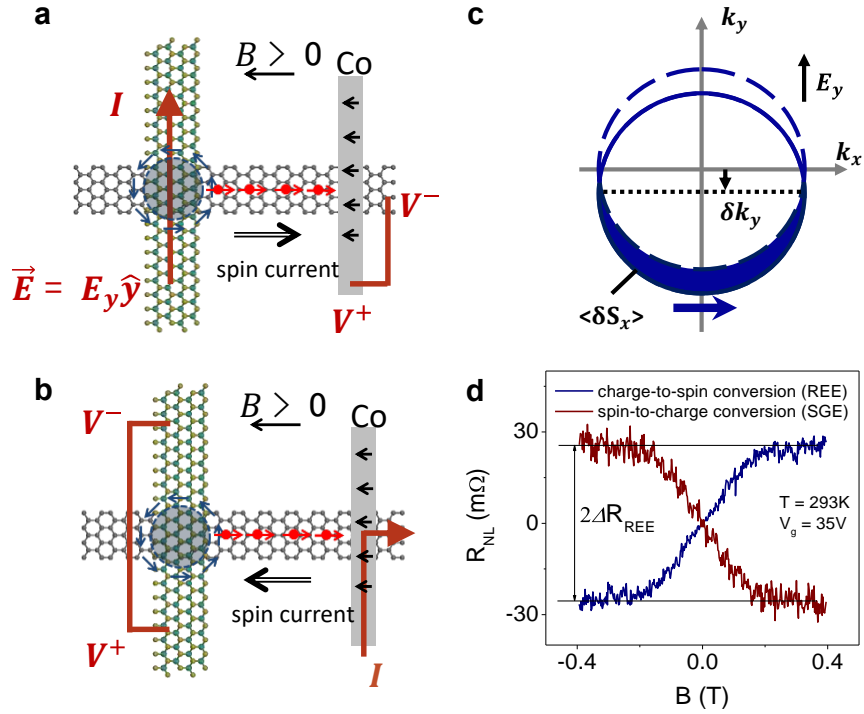

**a.** REE measurement protocol: a current $I$ is injected along 2H-TaS$_2$ and the nonlocal voltage generated between the ferromagnetic contact and 5LG ($\Delta V_{NL} = V_+ - V_-$) is measured while sweeping the magnetic field $B = -B\hat{x}$. **b.** SGE measurement protocol: a current is injected at FM3, and the voltage drop across the 2H-TaS$_2$/5LG strip is recorded. **c.** Schematic illustration of spin-helical Rashba sub-band and associated REE mechanism. Dashed and solid circles represent, respectively, the Fermi surface of the Rashba sub-band before and after application of a charge current (electric field) along +y direction. The arrows winding around the circle represent the sub-band equilibrium spin polarization vector $s_\mathbf{k}$. Spin-momentum locking ($s_\mathbf{k} \cdot \mathbf{k} = 0$) generates net nonequilibrium spin polarization $\delta S_x$ due to applied current. **d.** Magnetic field dependence of REE and SGE nonlocal resistances at $T$ = 293 K and $V_g$ = 35 V. The measured REE and SGE signals are opposite to each other and antisymmetric in the magnetic field, in accordance with the general Onsager reciprocal relations. The signals are sensitive to the current-induced spin-polarization generated by all spin-split sub-bands in the vicinity of the Fermi level.

, the in-plane field was removed, and Hanle-type spin precession measurements were performed with a perpendicular field ($B_z$). Figure 1f shows that the injected nonequilibrium spin polarization undergoes in-plane precession about $B_z$, which results in a modulation of the nonlocal resistance. By fitting Hanle curves at different $V_g$ to a 1D Bloch model, we have extracted a spin diffusion length $\lambda_G$ in the range 2-3 µm and polarization P = 4.8% at room temperature (see SI).

Next, we carry out spin-sensitive electrical measurements to detect spin-charge conversion effects. We first discuss the REE setup. The nonlocal detection scheme employed to investigate charge-to-spin conversion[44,45] is depicted in Fig. 2a. An applied charge current $I$ in the 2H-TaS$_2$ strip generates a nonlocal voltage $V_{NL}$ between the Co contact (FM3) and the normal Pd contact on 5LG (NM2). To enable detection of nonlocal signals that originate from REE, a magnetic field $\boldsymbol{B} = -B\hat{\boldsymbol{x}}$ is applied to tilt the FM3 detector magnetization towards its hard axis direction. The presence of a driving electric field ($E_y$, along the $+\hat{\boldsymbol{y}}$ direction) shifts the spin-split Fermi surface of interfacial states, thereby producing an excess spin polarization density $\delta S_x$ with spin-moment pointing along the $\hat{\boldsymbol{x}}$ axis (see Fig. 2b). This nonequilibrium spin polarization diffuses away from the heterojunction and is detected by FM3. The local change of spin-dependent electrochemical potential is proportional to the projection of the nonequilibrium spin polarization density onto the magnetization direction of FM3. In the standard 1D channel approximation, the spin accumulation $\mu_s = \mu_\uparrow - \mu_\downarrow$ ($\uparrow,\downarrow = \pm\hat{\boldsymbol{x}}$) detected at the FM3 contact ($x = L \approx 1.5$ µm) reads as

$$\Delta\mu_s(V_g, B) \simeq \Delta\mu_{\text{REE}}(V_g)\sin(\theta)\, e^{-\frac{L}{\lambda_G(V_g)}}, \qquad (2)$$

where $\Delta\mu_{\text{REE}}$ is the current-induced spin accumulation at the heterojunction ($x = 0$) and $\theta$ is the FM3 magnetization angle with respect to the easy axis (see SI).

Figure 2c shows the nonlocal REE resistance

$$R_{\text{REE}} \equiv \frac{V_{\text{NL}}}{I} = -P\frac{\Delta\mu_s}{2|e|I}, \qquad (3)$$

measured at room temperature in the high carrier density regime with $V_g = +35$ V. The applied field $B$ is swept between $-0.4$ T and $0.4$ T. We observe an antisymmetric response $R_{\text{REE}}(-B) \simeq -R_{\text{REE}}(B)$ characterized by a linear behavior at small magnetic fields, followed by saturation on the scale of $B_{\text{sat}} \approx 0.2$ T (SI). The nonlocal signal accurately follows the relation $R_{\text{REE}} \propto \sin\theta$, which is a *key signature* of charge-to-spin conversion via REE (Eq. 2). It is important to note that stray fields localized near the edge of ferromagnetic contacts can generate an ordinary Hall effect (OHE)[46], with the symmetry $R_{\text{OHE}}(B) = -R_{\text{OHE}}(-B)$, thus mimicking a REE signal. To quantify the spurious contribution to our measured signal, we performed micromagnetic simulations to evaluate the stray field profile and the induced OHE voltage between FM3 and NM2. Our calculations, under very conservative assumptions, give $R_{\text{OHE}} < 2.1$ m$\Omega \ll R_{\text{REE}} \approx 30$ m$\Omega$ for an applied field of 0.4 T at $V_g = +35$. The calculated upper bound (which disregards screening effects further reducing $R_{\text{OHE}}$) is far too weak to explain the data. Furthermore, the observed monotonic increase of $R_{\text{REE}}$ with $V_g$ is opposite to the typical behavior of OHE, which instead would result in a pronounced decrease of the nonlocal signal with $V_g$ in the high-density regime. Another important evidence is the saturation field scale $R_{\text{REE}} \approx 0.2$ T observed in the REE measurement and control spin-valve measurements, much lower than the OHE saturation field scale (estimated as $B_z \approx 0.5$ T); see SI for further details.

**Figure 3| Gate-voltage characteristics and charge-to-spin conversion efficiency**

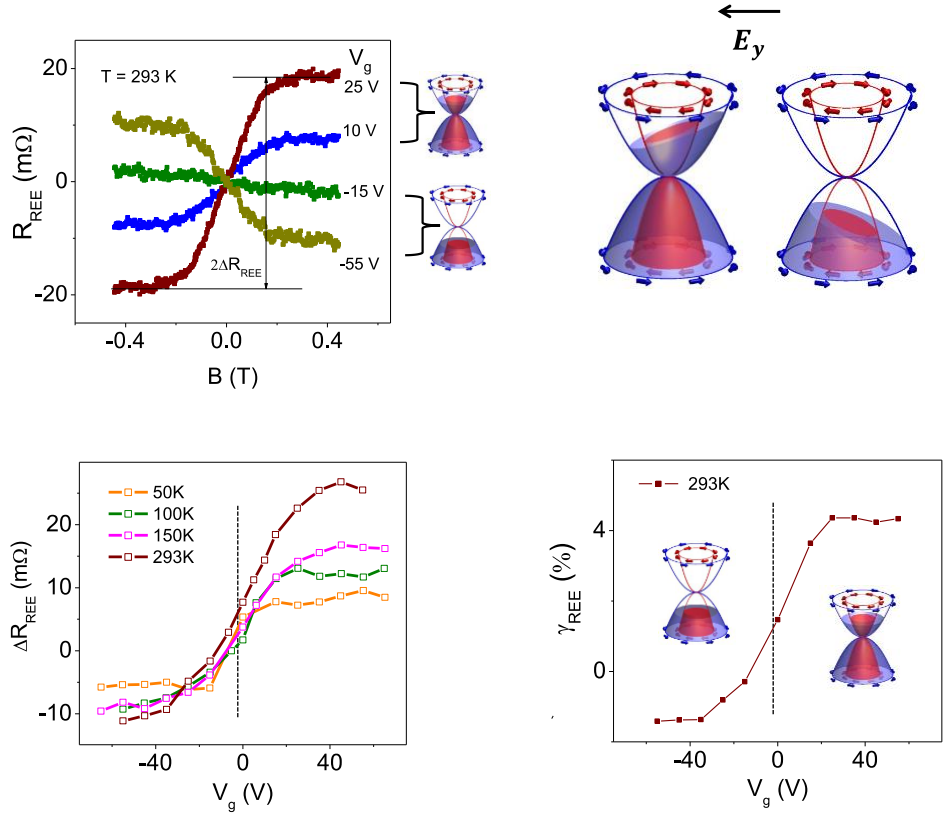

**a**. Nonlocal resistance $R_{\text{REE}}$ at room temperature as function of applied magnetic field $\boldsymbol{B} = -B\hat{\boldsymbol{x}}$ at selected values of gate voltage $V_g$. The signal vanishes near the CNP ($V_g \approx$ -3 V) for all values of the applied field $B$. The $R_{\text{REE}}$-sign is determined by the charge carrier polarity according to $\text{sign}\, R_{\text{REE}} = p\, \text{sign}\, B$, where $p = 1$ for electrons ($V_g > 0$) and $p = -1$ for holes ($V_g < 0$), as expected for REE originating from spin-helical 2D Dirac states. The $\boldsymbol{B}$-field modulation of the nonlocal signal $R_{\text{REE}}(B)$ follows accurately the relation $R_{\text{REE}} \propto \sin\theta$ for all $V_g$, demonstrating that it results from diffusive $\hat{\boldsymbol{x}}$-spin polarized currents generated at the heterojunction. **b**. Schematic illustration of the REE-induced spin polarization imbalance under an external electric field. The blue and red sub-bands represent schematic spin-split 2D Dirac states of a graphene-based heterostructure, having opposite spin helicity. Interfacial broken inversion symmetry endows spin-split states with counter-rotating spin textures. The Fermi level lies in the conduction band (CB-left) [valence band (VB-right)]. The applied electric field $E_y$ modifies the occupation probability of spin-helical electrons in $k$ space, which results simultaneously in a charge current ($J_y$) and a nonequilibrium spin polarization density with spin moments along the $\hat{x}$ direction ($\delta S_x$) (cf. Fig. 2b). The blue and red filling of the 2D Dirac cones represent the occupied states in applied electric field for majority (counterclockwise spins) and minority (clockwise spins) sub-bands. **c.** $R_{REE}$ as a function of gate voltage at selected temperatures. **d.** Lower-bound to charge-to-spin conversion efficiency $\gamma_{REE}$ as a function of gate voltage at room temperature.

The Onsager reciprocal experiment (the SGE detection) is shown in Fig. 2b. Here, a charge current is injected from FM3 to NM2 to generate a finite spin-dependent electro-chemical potential at the FM3/5LG interface in the presence of an external field $\boldsymbol{B} = -B\hat{x}$. The resulting $x$-polarized spin current $J_x^x = \sigma_s \nabla_x (\mu_\uparrow - \mu_\downarrow)$ diffuses through the graphene channel and is converted to a transverse charge current at the heterojunction, $J_y = \sigma_c \nabla_y (\mu_\uparrow + \mu_\downarrow)$, where $\sigma_{s(c)}$ is spin-conductivity of 5LG (charge conductivity of 5LG/2H-TaS$_2$). This near-equilibrium spin-charge conversion process is detected as an open-circuit voltage across the 2H-TaS$_2$ strip. Figure 2d shows the $B_x$ dependence of the room-temperature nonlocal resistance $R_{SGE}$ measured at $V_g = +35$ V. The direct (REE) and inverse (SGE) data are seen to satisfy the Onsager reciprocal relation, $R_{SGE}(B) = R_{REE}(-B)$, within experimental accuracy. Onsager reciprocity in SOC transport phenomena has been reported previously for SHE/inverse SHE (ISHE) measurement in metals[47] and, recently, for REE/SGE in metallic bilayers[48].

From a symmetry standpoint, the Onsager-reciprocal nonlocal transport in our device could originate from **(i)** interfacial REE/SGE in 5LG/2H-TaS$_2$ and **(ii)** bulk SHE/ISHE within bulk 2H-TaS$_2$. To determine which mechanism is responsible for the charge-spin interconversion of $x$-spin-polarized carriers, we tune the carrier density in 5LG by the application of a gate voltage. The spin texture of the interfacial states and the associated REE mechanism for generation of $x$-spin-polarized electrons when Fermi level lies in the conduction/valence band is illustrated in Fig. 3b.

A strong gate-tunability, with a clear antisymmetric behavior $R_{REE} \propto \sin\theta$ for all $V_g$ is observed (see Figs. 3a,3c). The most prominent feature is the sign change in $R_{REE}$ as $V_g$ is tuned across the CNP of the heterostructure, which rules out a bulk effect within the 2H-TaS$_2$ thin film, such as SHE. It is interesting to note that $R_{REE}$ increases with temperature at a given gate voltage. This is because that the REE efficiency (see Eq. (4) below) is related to the 5LG resistivity and the current flows in the 5LG at the junction and these parameters are temperature dependent. The observed ambipolar character of $R_{REE}(V_g)$, which vanishes near the CNP, and the monotonic increase of $R_{REE}$ with $|V_g|$ are unambiguous signatures of REE.

It is instructive to contrast our findings to recent reports of spin-charge conversion in graphene on group VI dichalcogenides[37,38,49]. In these works, the TMD is either a semiconductor TMD[37,38] or a semimetal WTe2[49]; measurements are performed in the absence of a gate control or with limited gate effect where gate reversal of the nonlocal resistance sign (the *key signature* of REE mediated by 2D Dirac fermions) is absent. In Ref.[38] the nonlocal resistance displays a steep decrease towards zero as $B_x$ approaches $B_{sat}$, indicating that $z$-polarized spins are converted to a charge signal *via* ISHE. In addition, a component linear with $B_x$ is present, and is attributed to bulk ISHE within the semiconducting TMD. However, the lack of gate tunability in Ref. [38] precludes the full discrimination between interfacial and bulk SOC effects. In Ref.[37] the gate tunability of in-plane (SGE) and out-of-plane (ISHE) spins is limited to low temperatures (and for electron-type carriers), indicating a fragile Rashba SOC[28]. Finally, the results in Ref.[49] clearly show that SHE within WTe2 is the active SOC transport mechanism; the nonlocal resistance is positive at all $V_g$, showing a maximum around the CNP due to increased spin absorption.

The lower bound to the REE efficiency $\gamma_{REE}$ is found as

$$\gamma_{REE} = \frac{V_{NL}}{\rho_{5LG}\, I_{5LG}} \frac{w_{TaS_2}}{P\, \lambda_G\, e^{-L/\lambda_G}}, \qquad (4)$$

where $I_{5LG}$ is the current flowing through 5LG (5LG is in the range of 18-60 $\mu$A depending on the applied gate voltage and temperature) and $w_{TaS_2}$ is the effective flake within which charge-spin conversion takes place. At room temperature, $\gamma_{REE}$ ranges from $-1.4\%$ at $V_g = -55$ V to $+4.3\%$ at $V_g = +55$ V (Fig. 3d). The sign change of $\gamma_{REE}$ originates from the sign reversal of $R_{REE}$ with $V_g$. This fine degree of electrical control over the nonlocal signal cannot be attributed to the spin transport characteristics of 5LG (*i.e.* the factor $\lambda_G\, e^{-L/\lambda_G}$) given the weak $V_g$ dependence of the spin relaxation length of graphene (see SI). Rather, it indicates that $R_{REE}$ is dominated by current-induced spin accumulation at the heterojunction (see the $\Delta\mu_{REE}(V_g)$ term in Eq. (2). The observed all-electrical spin-switching effect, *i.e.* sign-reversal of $R_{REE}$ with a gate voltage, is a manifestation of emergent spin-helical 2D Dirac fermions[28]. These experimental results agree qualitatively well with the model calculations as described in what follows.

To elucidate the nature of interfacial spin-orbit interactions in the van der Waals device, we carried out relativistic electronic structure calculations for a representative 5LG/bilayer-2H-TaS$_2$ supercell. Figures 4a and b show the crystal structure and band structure near the Dirac point (*K*) of 5LG/2H-TaS$_2$, respectively. The spin-splitting of 5LG Dirac-like states is clearly observed,

**Figure 4 | Relativistic electronic structure and tight-binding transport calculations.**

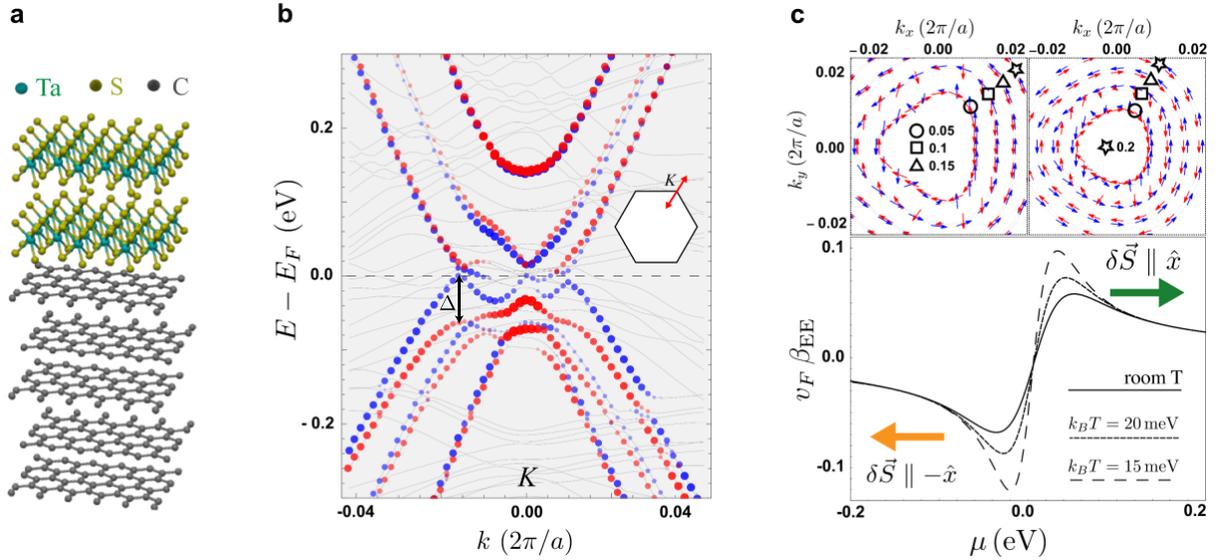

**a**. Supercell of the heterostructure built from Bernal stacked 5LG and bilayer 2H-TaS$_2$. **b**. Energy bands near the vicinity of the $K$ point (the wavevector path shown in the inset). $a = 0.14$ nm is the lattice scaling of graphene. Spin-split states prominently localized on carbon atoms in 5LG are indicated by red and blue dots. The maximum spin splitting near the Fermi level is on the order of 30 meV (electron sector) and 70 meV (hole sector, indicated by the double arrow). 2H-TaS$_2$ bands are shown in background. **c**. Top panel shows the spin-split Fermi surface and spin polarization texture of electronic states at selected energies ($E = 0.05, 0.1, 0.15,$ and $0.20$ eV) obtained from a minimal tight-binding model of pristine 5LG with proximity-induced SOC (see SI). To aid visualization, only 2 pairs of spin-split states are shown. Spin-split states mapped out in the left/right panel correspond to the electron states lying closest/second closest to the Dirac point as shown in (b). The lower panel shows the dimensionless microscopic figure of merit $v_F \beta_{EE}$ of the interfacial carbon layer as function of chemical potential at selected temperatures obtained from the minimal tight-binding model ($v_F = 10^6$ m/s is the Fermi velocity of massless 2D Dirac fermions).

with spin-gap reaching values as large as 70 meV for holes. The Fermi level shift, with respect to the unperturbed Dirac point, is only about +20 meV, despite the strong interaction between 5LG and 2H-TaS$_2$ (particularly visible in the hole bands). To estimate the efficiency of charge-to-spin conversion by spin-split states in 5LG, we carry out tight-binding transport simulations informed by the ab initio electronic structure (SI). The Fermi surface of a pristine 5LG with proximity-induced SOC is mapped out in Fig. 4c. Owing to spin-lattice-pseudospin entanglement[27,28,50], the magnitude $|s_\mathbf{k}|$ is sensitive to the Fermi wavevector. This is in contrast to conventional Rashba-split 2D electron gases[30,31], where the polarization is maximal near the Fermi level ($|s_\mathbf{k}| = \hbar/2$). In 5LG, the **k**-resolved spin texture displays a rich evolution from the CNP, where SOC trigonal warping effects dominate, to the highest accessible energies in the experiment ($|E_F| \approx 0.10 - 0.15$ eV), where well-established counter-rotating spin textures emerge.

The microscopic figure of merit $\beta_{EE}$ (units s/m) obtained from the tight-binding model is shown in the Fig. 4c (bottom panel). $\beta_{EE}$ increases monotonically from zero at the CNP to its maximum allowed value ($|\beta_{EE}| \approx 0.1/v_F$) when the chemical potential $|\mu|$ is on the order of the average spin gap $\Delta = \langle \Delta_\mathbf{k} \rangle$ (see SI for further details). The ambipolar effect—with the $\beta_{EE}$-sign determined by the charge carrier polarity—is a direct manifestation of spin-pseudospin coupling in an inversion-asymmetric 2D material[28]. The spin polarization of majority Rashba sub-bands (highlighted in blue in Fig. 4c) rotates anticlockwise, which explains the observed positive (negative) REE sign for electrons (holes) (see Fig. 3c). This ambipolar effect is in contrast to charge-to-spin conversion by surface states of topological insulators[36], where nonequilibrium spins are oriented along the same direction for both *n*- and *p*-type carriers. The slow decay of $\beta_{EE}$ in the high-density regime ($|\mu| \gtrsim \Delta$) signals the onset of minority- (clockwise) and majority-spin (anticlockwise) Rashba sub-bands with large polarization, $|s_\mathbf{k}| \cong \hbar/2$. This behavior is little sensitive to the relaxation time $\tau_{\vec{k}}$ (e.g. due to static disorder[28]), and thus provides a useful transport fingerprint of spin helical states.

The microscopic figure of merit $\beta_{EE}$ is linked to the charge-to-spin conversion efficiency $\gamma_{EE}$ accessible in the experiment through the relation $\gamma_{EE} = \vartheta_G \times \beta_{EE}$, where $\vartheta_G \propto \sigma_s/\lambda_G$ is the spin channel efficiency (SI). The tendency for signal saturation at high gate voltage in Fig. 3c (c.f. Fig. 3a) is thus reasonably explained by the combined effect of a slow decay in $\beta_{EE}$ in the high-density regime (Fig. 4c) and a moderate variation of the 5LG spin channel efficiency within the same Vg range, which is more pronounced in the hole sector. Thermal carrier activation plays an important role in the device operation,[28] especially at low chemical potential $|\mu| \lesssim \Delta$, where the energy- and wavevector-dependence of the spin texture are the most prominent (Fig. 4c). As such, for low $V_g$, thermally excited carriers experience less in-plane spin polarization due to a swift variation of $|s_\mathbf{k}|$ with Fermi wavevector (Fig. 4c), resulting in a decrease of REE efficiency with increasing temperature. This feature can also be inferred from the low-temperature behavior of $\beta_{EE}$, which shows a swift variation around the CNP (long-dashed curve in Fig. 4c, bottom panel). On the contrary, at high chemical potential, the spin texture in the vicinity of the Fermi level is well-established, i.e. $|s_\mathbf{k}|| \cong \hbar/2$ for $|\mu| \gtrsim \Delta$. Thus, in the high-density regime the REE efficiency is expected to display a moderate or even weak temperature dependence, consistent with the solution of Boltzmann transport equations (see Fig. 4c and SI).

In summary, we observed room-temperature current-induced spin polarization (REE) and its Onsager reciprocal phenomenon (SGE) in a 5LG/2H-TaS$_2$ heterostructure. The device nonlocal resistance can be controlled effectively by a gate voltage, with sign reversal as the Fermi level crosses the charge neutrality point. At room temperature, the lower bound of the charge-to-spin conversion efficiency ranges from -1.4 % for hole carriers, the ON state with excess spin "↓" to +4.3% for electron carriers, the "ON" state with excess spin "↑". The OFF state with zero net spin polarization is achieved at charge neutrality. The sign reversal observed in this work originates from the formation of spin-helical 2D Dirac-like fermions residing at the interface of 5LG/2H-TaS$_2$. The all-electrical spin-switching effect, *i.e.* the reversal of nonequilibrium spin polarization vector by the application of a gate voltage, demonstrated in this work opens doors to energy-efficient generation and manipulation of spin currents using nonmagnetic van der Waals materials.

**Note:** After completion of the current work, two experimental works were announced reporting the observation of gate-tunable room-temperature SGE in graphene/TMD heterostructures.[51,52] Ref. [52] reports the observation of interfacial SGE and ISHE in monolayer-graphene/WS$_2$, in accord with theoretical predictions for van der Waals bilayers where spin-valley coupling and Rashba SOC are concurrent.[23,28] Ref. [51] reports interfacial SGE in a heterostructure of CVD graphene and 1T'-MoTe$_2$, a metallic (group-VI) TMD. Similar to our work, the research[51,52] demonstrates the spin-switching effect with SGE resistance changing sign across the CNP[28]. However, unlike this work, REE measurements are not reported.

**Methods**

The heterostructure was fabricated by vertical assembly of exfoliated 2H-TaS$_2$ and few-layer graphene on heavily doped Si wafer topped with a layer of 285 nm thickness of SiO$_2$. The heavily doped Si was used as the gate. The bulk 2H-TaS$_2$ was grown by chemical vapor transport method using iodine as the transport agent. We use commercial Kish graphite for the source material of graphene. The thickness of 2H-TaS$_2$ was measured *via* atomic force microscopy. The number of layers and stacking sequence of the graphene flake were determined by Raman spectroscopy. Electrodes were patterned by e-beam lithography and e-beam evaporation of metals in a vacuum chamber of base pressure $10^{-7}$ Torr. Electrical measurements were performed using a standard low-frequency lock-in technique in a home-built cryostat with rotatable sample stage and a room-temperature magnet of maximum field 0.8 T.

Density functional theory calculations of electronic structure of graphene multilayers on TaS$_2$ are carried out with employment of Plane-wave/pseudopotential, Perdew-Burke-Ernzerhof (PBE), exchange-correlation energy functional, projector-augmented wave pseudopotentials.[53,34] The unit cells are chosen to consist of a 3 × 3 TaS$_2$ supercell and a 4 × 4 graphene supercell resulting in a small interlayer mismatch of less than 2 percent. The first Brillouin zone of the heterostructure was sampled using a 331 k-point grid. Van der Waals interactions were included using the opt88 functional[55]. All structures were fully relaxed until the force on each atom was less than 0.01 eV / Å. To visualize the Fermi surfaces, we interpolated the PBE band structure using maximally localized Wannier functions[56,57].

Tight-binding transport calculations are carried out to determine the spin-charge conversion characteristics of 5LG/2H-TaS$_2$. The graphene multilayer with ABABA stacking was described by the standard Slonczewski-Weiss-McClure model[58] supplemented with interface-induced Rashba interaction on the two carbon layers closest to 2H-TaS$_2$. The tight-binding parameters are adjusted until the spectrum qualitatively reproduces the *ab initio* spectrum of states predominantly localized on 5LG. Boltzmann transport equations incorporating electron-impurity scattering processes within the standard relaxation time approximation are used for the calculation of $\beta_{EE}$.

## Acknowledgments

S.C. and L.L. are grateful for valuable discussions with Y. Otani on Onsager reciprocity in REE experiments. A.F., S.A.C. and S.C. are indebted to B. van Wees' group for discussions on the impact of stray fields in spin transport measurements. We would particularly like to thank A. Kaverzin for sharing finite-element simulations of the leakage current in the REE geometry. We acknowledge support from Korea NRF (Grant No. 2019M3F3A1A03079760) (S.C.), Korea NRF (Grant No. 2017R1D1A1B03030877) (L.L.), the Institute for Basic Science under Grant No. IBS-R009-G2 (B.S.K and C.K.), the Royal Society of London through a University Research Fellowship (A.F.) and a Global Challenges Research Fund award under Grant No. CHG-R1-170063 (J.Z. and J.L.).

## Competing interests

The authors declare no competing financial interests.

# References

1. I. Zutic, J. Fabian, S. Das Sarma. Spintronics: Fundamentals and Applications. *Rev. Mod. Phys*. **76**, 323 (2004)

2. A. Soumyanarayanan, N. Reyren, A. Fert, C. Panagopoulos, Emergent phenomena induced by spin-orbit coupling at surfaces and interfaces. *Nature* **539**, 509 (2016).

3. J. Sinova, S. O. Valenzuela, J. Wunderlich, C. H. Back, T. Jungwirth, Spin Hall effects. *Rev. Mod. Phys*. **87**, 1213 (2015).

4. W. Han, R. K. Kawakami, M. Gmitra, J. Fabian, Graphene Spintronics. *Nat. Nanotechnol.* **9**, 794 (2014).

5. N. Tombros, C. Jozsa, M. Popinciuc, J. T. Jonkman, B. J. van Wees, Electronic spin transport and spin precession in single graphene layers at room temperature. *Nature* **448**, 571 (2007).

6. M. Drogeler, F. Volmer, M. Wolter, B. Terres, K. Wantanabe, T. Taniguchi, G. Guntherodt, C. Stampfer, B. Beschoten, Nanosecond Spin Lifetimes in Single- and Few-Layer Graphene–hBN Heterostructures at Room Temperature. *Nano Lett.* **14**, 6050 (2014).

7. J. Xu, T. Zhu, Y. K. Luo, Y-M. Lu, R. K. Kawakami, Strong and Tunable Spin-Lifetime Anisotropy in Dual-Gated Bilayer Graphene. *Phys. Rev. Lett.* **121**, 127703 (2018).

8. J. C. Leutenantsmeyer, J. Ingla-Aynes, F. Jaroslav, B. J. van Wees, Observation of Spin-Valley-Coupling-Induced Large Spin-Lifetime Anisotropy in Bilayer Graphene. *Phys. Rev. Lett.* **121**, 127702 (2018).

9. Q. H. Wang, K. Kalantar-Zadeh, A. Kis, J. N. Coleman, M. S. Strano, Electronics and optoelectronics of two-dimensional transition metal dichalcogenides. *Nature Nanotechnology* **7**, 699 (2012).

10. A. Avsar, D. Unuchek, J. Liu, O. L. Sanchez, K. Watanabe, T. Taniguchi, B. Ozyimaz, A. Kis, Optospintronics in Graphene via Proximity Coupling. *ACS Nano* **11**, 11678 (2017).

11. Y. K. Luo, J. Xu, T. Zhu, G. Wu, E. J. Macormick, W. Zhan, R. Neupane, R. K. Kawakami, Opto-valleytronic spin injection in monolayer $MoS_2$/few-Layer graphene hybrid spin valves. *Nano Lett.* **17**, 3877 (2017).

12. J.H. Garcia, M. Vila, A.W. Cummings, S. Roche. Spin Transport in Graphene/Transition Metal Dichalcogenide Hetero- structures. *Chem. Soc. Rev*. **47**, 3359 (2018).

13. A. Avsar, H. Ochoa, F. Guinea, B. Ozyilmaz, B. J. van Wees, I. J. Vera-Marun. Colloquium: Spintronics in Graphene and Other Two-Dimensional Materials. arXiv 1909.09188 (2020)

14. A. Avsar, J. Y. Tan, T. Taychatanapat, J. Balakrishnan, G. K. W. Koon, Y. Yeo, J. Lahiri, A. Carvalho, A. S. odin, E. C. T. O'Farrell, G. Eda, A. H. Castro Neto, B. Ozyilmaz, Spin-orbit proximity effect in graphene. *Nature Commun*. **5**, 4875 (2014).

15. Z. Wang, D-K. Ki, H. Chen, H. Berger, A. H. MacDonald, A. F. Morpurgo, Strong interface-induced spin–orbit interaction in graphene on $WS_2$. *Nat. Commun*. **6**, 8339 (2015).

16. Z. Wang, D-K. Ki, J. Y. Khoo, D. Mauro, H. Berger, L. S. Levitov, A. F. Morpurgo, Origin and magnitude of 'designer' spin-orbit interaction in graphene on semiconducting transition metal dichalcogenides. *Phys. Rev. X* **6**, 041020 (2016).

17. J. Sichau, M. Prada, T. Anlauf, T. J. Lyon, B. Bosnjak, L. Tiemann, R. H. Blick, Resonance microwave measurements of an intrinsic spin-orbit coupling gap in graphene: A possible indication of a topological state. *Phys. Rev. Lett.* **122**, 046403 (2019).

18. T. S. Ghiasi, J. Ingla-Aynes, A. A. Kaverzin, B. J. van Wees, Large proximity-induced spin lifetime anisotropy in transition-metal dichalcogenide/graphene heterostructures. *Nano Lett.* **17**, 7528 (2017).

19. L. A. Benítez, J. F. Sierra, W. Savero Torres, A. Arrighi, F. Bonell, M. V. Costache, S. O. Valenzuela, Strongly anisotropic spin relaxation in graphene-transition metal dichalcogenide heterostructures at room temperature. *Nat. Phys.* **14**, 303 (2018).

20. A. Ferreira, T.G. Rappoport, M.A. Cazalilla, A.H.C. Neto, Extrinsic Spin Hall Effect Induced by Resonant Skew Scattering in Graphene. *Phys. Rev. Lett.* **112**, 066601 (2014).

21. M. Milletari, A. Ferreira. Quantum Diagrammatic Theory of The Extrinsic Spin Hall Effect in Graphene. *Phys. Rev. B* **94**, 134202 (2016).

22. C. Huang, Y.D. Chong, M.A. Cazalilla, Direct Coupling between Charge Current and Spin Polarization by Extrinsic Mechanisms in Graphene. *Phys. Rev. B*. **94**, 085414 (2016).

23. M. Milletari, M. Offidani, A. Ferreira, R. Raimondi, Covariant Conservation Laws and the Spin Hall Effect in Dirac-Rashba Systems. *Phys. Rev. Lett.* **119**, 246801 (2017).

24. J.H. Garcia, M. Vila, A.W. Cummings, S. Roche. Spin Transport in Graphene/Transition Metal Dichalcogenide Hetero- structures. *Chem. Soc. Rev*. **47**, 3359 (2018).

25. Y.A. Bychkov, E.I. Rashba, Properties of a 2DEG with Lifted Spectral Degeneracy. *JETP Lett.* **39**, 78 (1984).

26. D. Kochan, S. Irmer, J. Fabian. Model Spin-Orbit Coupling Hamiltonians For Graphene Systems. *Phys. Rev. B* **95**, 165415 (2017).

27. E. I. Rashba, Graphene with structure-induced spin-orbit coupling: spin-polarized states,

spin zero modes, and quantum Hall effect. *Phys. Rev.* B **79**, 161409(R) (2009).

28. M. Offidani, M. Milletari, R. Raimondi, A. Ferreira, Optimal charge-to-spin conversion in graphene on transition-metal dichalcogenides. *Phys. Rev. Lett.* **119**, 196801 (2017).

29. V.M. Edelstein, Spin Polarization of Conduction Electrons Induced by Electric Current in Two-Dimensional Asymmetric Electron Systems. *Solid State Commun*. **73**, 233 (1990).

30. K. Shen, G. Vignale, R. Raimondi, Microscopic Theory of the Inverse Edelstein Effect. *Phys. Rev. Lett.* **112**, 096601 (2014).

31. S. D. Ganichev, *et al.*, Spin-Galvanic Effect. *Nature* **417**, 153 (2002).

32. J. C. R. Sanchez, L. Villa, G. Desfonds, S. Gambarelli, J. P. Attane, J. M. de Teresa, C. Magen, A. Fert, Spin-to-charge conversion using Rashba coupling at the interface between non-magnetic materials. *Nat. Commun.* **4**, 2944 (2013).

33. H. Nakayama, Y. Kanno, H. An, T. Tashiro, S. Haku, A. Nomura, K. Ando, Rashba-Edelstein magnetoresistance in metallic heterostructures. *Phys. Rev Lett.* **117**, 116602 (2016).

34. Q. Song, H. Zhang, T. Su, W. Yuan, Y. Chen, W. Xing, J. Shi, J. Sun, W. Han, Observation of inverse Edelstein effect in Rashba-split 2DEG between SrTiO3 and LaAlO3 at room temperature. *Sci. Adv.* **3**, e1602312 (2017).

35. E. Lesne, Y. Fu, S. Oyarzun, J. C. Rojas-Sanchez, D. C. Vaz, H. Naganuma, G. Sicoli, J.-P. Attane, M. Janet, E. Jacquet, J.-M. George, A. Barthelemy, H. Jaffres, A. Fert, M. Bibles, L. Vila, Highly efficient and tunable spin-to-charge conversion through Rashba coupling at oxide interfaces, *Nat. Mater.* **15**, 1261 (2016).

36. K. Kondou, R. Yoshimi, A. Tsukazaki, Y. Fukuma, J. Matsuno, K. S. Takahashi, M. Kawasaki, Y. Tokura, Y. Otani, Fermi-level-dependent charge-to-spin current conversion by Dirac surface states of topological insulators. *Nat. Phys.* **12**, 1027 (2016).

37. T.S. Ghiasi, A.A. Kaverzin, P.J. Blah, B.J. van Wees. Charge-to-Spin Conversion by the Rashba-Edelstein Effect in Two- Dimensional van der Waals Heterostructures up to Room Temperature. *Nano Lett.* **19**, 5959 (2019).

38. C.K. Safeer *et al*, F. Room-Temperature Spin Hall Effect in Graphene/MoS2 van der Waals Heterostructures. *Nano Lett.,* **19**, 1074 (2019).

39. T. Shimada, F. S. Ohuchi, B. A. Parkinson, Work function and photothreshold of layered metal dichalcogenides. *Jpn. J. Appl. Phys*. **33**, 2696 (1994).

40. G. Giovannetti, P. A. Khomyakov, G. Brocks, V. M. Karpan, J. van den Brink, P. J. Kelly, Doping graphene with metal contacts. *Phy. Rev. Lett*. **101,** 026803 (2008).

41. T. Maassen, F. K. Dejene, M. H. D. Guimaraes, C. Jozsa, B. J. van Wees, Comparison between charge and spin transport in few-layer graphene. *Phys. Rev. B* **83**, 115410 (2011).

42. S. Latil, L. Henrard, Charge Carriers in Few-Layer Graphene Films. *Phys. Rev. Lett.* **97**, 036803 (2006).

43. W. Bao *et al.*, Stacking-Dependent Band Gap and Quantum Transport in Trilayer Graphene. *Nat. Phys.* **7**, 948 (2011).

44. L. Vila, T. Kimura, Y. Otani, Evolution of the spin Hall effect in Pt nanowires: size and temperature effects. *Phys. Rev. Lett*. **99**, 226604 (2007).

45. S.O. Valenzuela. Electrical Detection of Spin Currents: The Spin-Current Induced Hall Effect. *J. Appl. Phys.* **101**, 09B103 (2007).

46. F.G. Monzon, D.S. Patterson, M.L. Roukes, Characterization of Individual Nanomagnets by the Local Hall Effect. *J. Magn. Magn. Mater.* **195**, 19 (1999).

47. T. Kimura, Y. Otani, T. Sato, S. Takahashi, S. Maekawa, Room-Temperature Reversible Spin Hall Effect. *Phys. Rev. Lett.* **98**, 156601 (2007).

48. H. Isshiki, P.Mudli, J. Kim, K. Kondou, Y. Otani, Experimentally Determined Correlation between Direct and Inverse Edelstein Effect in Bi2O3/Cu Interface by Means of Spin Absorption Method Using Lateral Spin Valve Structure. arXiv 1901. 03095 (2019)

49. B. Zhao, D. Khokhriakov, Y. Zhang, H. Fu, B. Karpiak, A. Md. Hoque, X. Xu, Y. Jiang, B. Yan, S. P. Dash, Observation of spin Hall effect in semimetal WTe2. arXiv:1812.02113 (2018).

50. D.V. Tuan, F. Ortmann, D. Soriano, S.O. Valenzuela, S. Roche, Pseudospin-Driven Spin Relaxation Mechanism in Graphene. *Nat. Phys.* **10**, 857 (2014).

51. A. M. Hoque *et al*. All-electrical creation and control of giant spin-galvanic effect in 1T-MoTe2/graphene heterostructures at room temperature. arXiv:1908.09367 (2019)

52. L. A. Benitez, *et al*. Room-Temperature Spin Galvanic and Spin Hall Effects in van der Waals Heterostructures. *Nat. Mater*. **19**, 170 (2020).

53. G. Kresse, J. Furthmuller, Efficient iterative schemes for ab initio total-energy calculations using a plane-wave basis set. *Phys. Rev. B* **54**, 11169 (1996).

54. J.P. Perdew. K. Burke, M. Ernzerhof, Generalized Gradient Approximation Made Simple. *Phys. Rev. Lett.* **77**, 3865 (1996).

55. J. Klimes, D.R. Bowler, A. Michaelides, Van der Waals Density Functionals Applied to

Solids. *Phys. Rev. B* **83**, 195131 (2011).

56. N. Marzari, D. Vanderbilt, Maximally Localized Generalized Wannier Functions for Composite Energy Bands. *Phys. Rev.* B **56**, 12847 (1997).

57. A.A. Mostofi *et al*. Wannier90: A Tool for Obtaining Maximally-Localised Wannier Functions. *Comput. Phys. Commun.* **178**, 685 (2008).

58. E. McCann, M. Koshino, The electronic properties of bilayer graphene. *Rep. Prog. Phys.* **76**, 056503 (2013).

# Supplemental Information

## 1. Characterization of thin film 2H-TaS$_2$

The quality of 2H-TaS$_2$ crystal is characterized by measuring the temperature-dependent electrical resistivity of an exfoliated TaS$_2$ thin layer of similar thickness used for REE. The resistivity is shown in Fig. S1. As indicated by the black arrow, one can identify the charge-density-wave-induced kink,[S1] which not only confirms the 2H phase of TaS$_2$ but also testifies to the high quality of the crystals used in the experiment. Resistivity of the layer changes from $22 \times 10^{-8}$ to $380 \times 10^{-8}$ $\Omega$m as temperature changes from 22 K to 293K. The layer thickness is 12 nm, and the 2D resistivity is 18 $\Omega$ at 22 K and 314 $\Omega$ at room temperature. At 50 K the 2D resistivity is 79 $\Omega$.

## 2. Device Fabrication

A five-layer graphene (5LG) and a thin layer of 2H-TaS$_2$ were mechanically exfoliated from Kish graphite and bulk crystal of 2H-TaS$_2$ respectively. Using a scotch tape, the layers were cleaved and deposited on 285 nm SiO$_2$/Si substrate. Highly doped Si was used as a back gate. By using a PC-film-covered PDMS stamp, the TaS$_2$ flake was picked up from the substrate at 60ºC and released on the 5LG at an elevated temperature. The PC film was washed out by chloroform followed by rinsing with acetone and IPA. The 5LG flake has a length of 35 µm and a width of 3.6 µm. The width of the TaS$_2$ flake at the heterostructure region is approximately 6.5 µm.

Electrodes were patterned by *e*-beam lithography and deposited by *e*-beam evaporation. Cr/Pd (3/45 nm) electrodes were used as normal Ohmic contacts for 2H-TaS$_2$ and 5LG. Before deposition of the ferromagnetic electrodes Co/Pd (50/5 nm), 0.6 nm-thick Ti was deposited and oxidized naturally to form TiO$_x$ tunnel barriers for enhanced spin injection and detection. 5 nm-thick Pd was used as a capping layer to avoid oxidation of the Co surface. From left to right of the 5LG (see Fig. 1b), we have made four Co electrode contacts with varying widths of 250, 300, 400, and 300 nm, respectively. Before wire bonding to the chip carrier, the sample surface was coated with PMMA to prevent the device from being contaminated by air exposure. At $V_g = 0$, the contact resistance $R_C$ at $T = 300$ K between Co electrode and 5LG is about 3 k$\Omega$. The quality of the contact for spin injection can be assessed by the ratio of the contact resistance $R_C$ to the spin resistance $R_l$

= $\rho\lambda_G/W_G$, here $\lambda_G$ is the spin diffusion length of 5LG and $W_G$ is the 5LG channel width. In this device $R_C/R_l \approx 10$, indicating fairly good quality of the ferromagnetic contacts.[S2-S4]

## 3. Atomic force microscopy (AFM) measurement of 2H-TaS$_2$ layer thickness

Tapping mode AFM measurement was performed to determine the thickness of 2H-TaS$_2$ layer in our device. AFM measurement shows that the thickness of TaS$_2$ layer is about 15 nm (Fig. S2).

## 4. Raman study: determination of number of layers and stacking sequence of graphene

The number of graphene layers and stacking sequence were determined by Raman spectroscopy using photon energy 2.33 eV (532 nm). A monolayer graphene was used as reference. We evaluated the ratio of peak intensities $I(\text{Si})/I(\text{Si}_0)$ with and without a graphene layer. Figure S3 shows the Raman G and 2D peak of the reference flake and few-layer graphene in our device. The $I(\text{Si})/I(\text{Si}_0)$ for monolayer was 0.900 while it was 0.654 for the few-layer graphene (Fig. S3 inset a). The number of layers was determined using the Si peak analysis method.[S5] Furthermore, from the shape of Raman 2D peak (Fig. S3 inset b), we identified the stacking sequence of the five-layer graphene (5LG) flake as ABABA (Bernal type).[S6]

## 5. TaS$_2$ thin film resistance as a function of gate voltage and temperature

Figures S4a and S4b show the gate voltage and temperature dependence of two-probe resistance of the 15 nm-thick TaS$_2$ in our device. The two-probe resistance of 2H-TaS$_2$, $R_{\text{TaS2}}$, remains constant under applied back-gate voltage. In the temperature range 50 K–300 K, where we investigated the charge-to-spin interconversion due to REE and its Onsager reciprocal, the temperature-dependent $R_{\text{TaS2}}$ shows metallic behavior. Below 27 K, the resistance of TaS$_2$ shows a sharp increase with decreasing temperature, signaling the formation of charge density wave.[S1] The spin transport properties in this regime is beyond the scope of our study.

## 6. Characterization of spin diffusion in 5LG

Figures 1e and 1f in main text show lateral spin valve and Hanle spin precession measurements with an applied out-of-plane magnetic field to characterize the diffusive spin transport in the 5LG area not covered by TaS$_2$. To extract the spin diffusion constant $D$ and spin relaxation time $\tau_s$, we used equation 3 in Ref.[S7]:

$$R_{SV} = \frac{P^2 \lambda_G \rho_{2D}}{W_G} \exp\left(-\frac{L}{\lambda_G}\right)(1 + \omega_L^2 \tau_s^2)^{-\frac{1}{4}}$$

$$\times \exp\left(-\frac{L}{\lambda_G}\left\{\sqrt{\frac{1}{2}\left(\sqrt{1 + \omega_L^2 \tau_s^2} + 1\right)} - 1\right\}\right)$$

$$\times \cos\left\{\frac{\tan^{-1}\omega_L \tau_s}{2} + \frac{L}{\lambda_G}\sqrt{\frac{1}{2}\left(\sqrt{1 + \omega_L^2 \tau_s^2} - 1\right)}\right\}. \quad \text{(Eq. S1)}$$

Here, $\omega_L = g\,\mu_B B$ is the Larmor frequency, $g = 2$ is the $g$ factor and $\mu_B$ is the Bohr magneton. The spin diffusion length is obtained from $\lambda_G = \sqrt{D\tau_s}$. At 293 K, $\lambda_G$ has weak gate-dependence ranging from 2 to 3 μm (Fig. S5a). The ferromagnetic contact polarization is $P = (G_\uparrow - G_\downarrow)/G = 4.8\%$, where $G_{\uparrow,\downarrow}$ is the spin dependent tunneling conductance between graphene and Co.

Figures S5b and S5c show spin precession curves with magnetic field applied in-plane (along the quasi-1D spin channel, $x$ axis) at $T = 293$ K and $T = 22$ K, respectively. By fitting the in-plane Hanle curve at $T = 293$ K, we obtained $\lambda_G = 2.19$ μm, which compares well with the value obtained by fitting the out-of-plane Hanle data. When a $B$ field of increasing magnitude is applied along the $+x$ direction, the magnetization of Co, initially aligned to the easy axis ($y$-axis), rotates and eventually becomes parallel to the $B$ field ($-x$ direction). The field dependence $\boldsymbol{M} = \boldsymbol{M}(\boldsymbol{B})$ was obtained from lineshapes for parallel and antiparallel configurations. Figure S5d shows that due to applied field along the $x$ axis, the Co magnetization tilts away from the easy axis towards $x$ axis and eventually saturates around B ≈ 0.2 T.

## 7. Micro-magnetic simulation of the FM3 stray field

We simulate the Co electrode as a rectangular planar structure of dimensions 400 nm × 5000 nm × 50 nm ($x$, $y$, $z$) with a rectangular cell size of 5 nm × 5 nm × 5 nm using the micro-magnetic package OOMMF, which implements the Landau–Lifshitz–Gilbert equation of motion for magnetic macrospins[S8] and includes the exchange interaction between the macrospins and the long ranged demagnetization fields. The parameters used as inputs into OOMMF are based on typical values for the saturation magnetization ($M_s$) and exchange coefficient ($A$) found in the literature for polycrystalline Co thin films: $M_s = 1.4$ MAm[-1] and $A = 3 \times 10$[-11] Jm[-1][S9]. The system was relaxed *via* the conjugate gradient minimizer method as the applied field in the $x$-direction was swept from 1000 mT to 0 mT in steps of 100 mT. The 3D magnetization vector map and the 3D demagnetization (stray) field vector map were generated for each magnetic field step in order to obtain the z component of the stray field ($B_z$) as a function of distance along the x direction from the Co electrode (labelled FM3 in Fig. 1a, main text).

Figure S6a shows the stray field profile $B_z(x,y)$ at $z = 0$ nm i.e. at interface between the Co electrode and 5LG for an applied field of 1000 mT in the negative x direction. $B_z$ is largest at the edges of the structure which are perpendicular to the magnetization vector. The profile $B_z(x)$, averaged along the central 2000 nm of the structure in the y-direction, is shown in Fig. S6b for two values of the applied field. As expected from the classical theory of fields generated by a magnetic dipoles[10]. $B_z$ falls away from the edge of the structure as $\approx x^{-3}$ - shown in more detail in Fig. S7. The inset to Fig. S7 shows the peak magnitude of $B_z(x,y)$ as a function of the applied field $B_x$. The stray field profile, $B_z(x,y)$ is used to calculate the size of the Ordinary Hall voltage $V_{\text{OHE}}$ generated between FM3 and NM2 (see Fig. S8).

## 8. Local ordinary Hall effect

In order to find $V_{\text{OHE}}$, we compute the electrostatic field $\varphi(x, y)$ in 5LG. We model the 2D flake as an infinite strip of width $W$ with dc-conductivity $\sigma_{xx}$ (in zero field); see Fig. S8. $\varphi(x, y)$ is computed from the solution of the Laplace equation with the boundary condition defined by the applied current $j_y(x, y = \pm W/2) = I\delta(x)$ in the REE experiment ($\delta$ is the Dirac-delta function). The solution reads as[S11]

$$\varphi(x, y) = -\frac{1}{\sigma_{xx}} \int_{-\infty}^{\infty} dk \, \frac{I \, e^{ikx}}{k \cosh\left(\frac{kW}{2}\right)} \sin(ky), \qquad \text{(Eq. S2)}$$

where we have neglected a parametrically small correction due to the nonzero off-diagonal term ($\sigma_{yx}(x, y)$) near the contact (FM3). This approximation is valid in the field range of interest, for which $|\sigma_{yx}| \ll \sigma_{xx}$.

The Hall field ($E_x$) between contacts FM3 and NM2 is determined by the requirement $j_x(x,y) = 0$. We find $E_x(x,y) \cong \rho_{xy}(x,y) j_y(x,y)$, where $j_y \cong -\sigma_{xx} \partial_y \varphi$ and $\rho_{xy}(x,y)$ is the 2D Hall resistivity induced by the $z$-component of the stray field in the 5LG basal plane ($z=0$):

$$\rho_{xy}(x,y) = -\frac{\sigma_{xy}(B_z(x,y))}{\sigma_{xx}(B_z(x,y))^2 + \sigma_{xy}(B_z(x,y))^2}. \tag{Eq. S3}$$

Explicit calculation gives

$$E_x(x,y) \simeq -\frac{\sigma_{xy}(B_z(x,y))}{\sigma_{xx}(0)^2} F(x/2W, y/2W), \tag{Eq. S4}$$

with $F(a,b) = \{\cot[\pi(i\,a + b + 1/4)] + 2\,\text{sech}[\pi(a + i\,b)] + \tan[\pi(i\,a + b + 1/4)]\}/4W$. The induced OHE voltage between the FM3 and NM2 contacts reads as

$$\Delta V(y) = -\int_{-x_{\text{FM3}}}^{x_{\text{NM2}}} dx\, E_x(x,y), \tag{Eq. S5}$$

where $x_{\text{NM2}}$ and $x_{\text{FM3}}$ are the positions of FM3 and NM2 contacts along the $x$-axis w.r.t. the current injection point, respectively. For the Hall conductivity of 5LG, we use a standard Drude model for degenerate electrons ($\sigma_{xx} = \sigma_0/(1 + \omega_c^2 \tau^2)$ and $\sigma_{xy} = -\sigma_0 \omega_c \tau/(1 + \omega_c^2 \tau^2)$),[S12] where $\sigma_0$ is the zero-field Ohmic conductivity, $\omega_c$ is the cyclotron frequency and $\tau$ is the transport time. In the experimentally relevant limit $1 \gg \omega_c^* \tau$, where $\hbar \omega_c^*$ is the typical cyclotron energy near the FM3 contact, we find (for electrons, $E_F > 0$):

$$R_{\text{OHE}} \equiv -\frac{\Delta V}{I} \cong R_0 \times \left(\frac{E_c \tau}{\hbar}\right) \times \Psi(y, W). \tag{Eq. S6}$$

In the above, $R_0 = 1/\sigma_0 \approx 220\,\Omega$ at room temperature in the high-density regime (Figure 1c, main text), $E_c = \hbar e/m^*$ is the typical cyclotron energy (per Tesla) and $\Psi$ is given by

$$\Psi(y, W) = \int_{-x_{\text{FM3}}}^{x_{\text{NM2}}} dx \, B_z(x, y) F(x/2W, y/2W) \,. \tag{Eq. S7}$$

The minus sign in the definition of $R_{\text{OHE}}$ (see Eq. S6) results from the convention $V_{\text{NL}} \equiv V_{\text{FM3}} - V_{\text{NM2}}$. The integral in Eq. S7 is performed numerically with the simulated stray field profile (Fig. S6). A reliable estimate for $E_c$ can be obtained from known effective masses for graphites[S13] and far-infrared magneto-optical measurements in few-layer graphene.[S14,S15] Here, we assume a "worst-case scenario" of a monolayer-like magneto-transport response with a long relaxation time, $E_c \tau / \hbar = 0.43$,[S12] obtained using $E_F = 0.1$ eV and $\hbar/\tau = 15$ meV.[S15, S16] The result is shown in Fig. S9. The average value of the OHE resistance across the strip width ($W \cong 3.6 \, \mu$m) is $\bar{R}_{\text{OHE}} \cong 2.1$ mΩ, which is much smaller than the detected nonlocal signal in the REE experiment[2]. It is worth noting that the non-local resistance REE, following the form of $M_x(B_x)$ rather than $M_z(B_x)$ (which shows saturation on qualitatively different field scales; Fig. S8), provides further evidence that the measured signal is not dominated by stray field induced OHE. Last but not least, the carrier density dependence of the OHE contribution (Eq. S6) has the typical scaling of a 2D electron gas (*i.e.*, $R_{\text{OHE}} \propto 1/n$), whereas the observed REE resistance increases with the electronic density for both electron and hole sectors.

## 9. Derivation of lower-bound Rashba-Edelstein effect (REE) efficiency

Due to REE induced in the 5LG, a current *I* flowing in TaS$_2$ induces a spin accumulation at the heterostructure region $\mu_s(0)$. This spin accumulation causes a 2D spin current $J_s(x)$ flowing into the 5LG spin channel. At a distance *x* from the heterostructure,

---

[2] In order to allow a simple algebraic treatment, our model assumes uniform conductivity ($\sigma_{xx}$) in the strip. On the other hand, the 5LG channel in the experiment is partly covered by 2H-TaS$_2$. The heterojunction is thus characterized by an Ohmic conductivity, $\sigma_{2D} \neq \sigma_{xx}$. From the measurements taken at room temperature in the high-density regime, we extract an upper bound of $\sigma_{xx}/\sigma_{2D} \lesssim 0.5$ (Figs. 1c-d; main text). The electrostatic field lines reaching the FM3 contact are thus much weaker than those of the simplified model (Eq. S2). Moreover, the FM3 contact itself provides extra screening. As such, our estimate $\bar{R}_{\text{OHE}}$ must be understood as a conservative upper bound for the OHE contribution in the experiment.

$$\mu_s(x) = \mu_s(0)\exp(-x/\lambda_G),\qquad(\text{Eq. S8})$$

$$J_s(x) = -\frac{\sigma_G}{e}\frac{\partial \mu_s}{\partial x} = \frac{\sigma_G \mu_s(0)}{e\lambda_G}\exp(-x/\lambda_G),\qquad(\text{Eq. S9})$$

where we assume that the spin conductivity equals the charge conductivity $\sigma_G$. Then,

$$J_s(0) = \frac{\sigma_G \mu_s(0)}{e\lambda_G}.\qquad(\text{Eq. S10})$$

The total charge current flows in TaS2, the interface, and 5LG. The interface current is the one that induces the spin accumulation by REE. However, it is difficult to quantify this current precisely. Instead, we use the known current flowing in 5LG, $I_{5LG}$, to compute a lower bound to the charge-to-spin conversion efficiency, $\gamma_{\text{REE}}$. This was done by treating the TaS2 and 5LG as two resistors in parallel and calculating the ratio $s$ of the current flowing in $I_{5LG}$ to total current $I$, $s = \frac{I_{5LG}}{I}$. Using $J_s(0) = \gamma_{\text{REE}} J_G = \gamma_{\text{REE}} Is/w_{\text{TaS}_2}$, where $w_{\text{TaS}_2}$ is the width of TaS2 at the junction, we find

$$\mu_s(0) = \frac{e\,\lambda_G\,\gamma_{\text{REE}}\,I\,s}{\sigma_G\,w_{\text{TaS}_2}}.\qquad(\text{Eq. S11})$$

At the Co electrode the measured non-local voltage is therefore

$$V_{\text{NL}} = \frac{P\mu_s(x)}{e} = \frac{P\mu_s(0)}{e}\exp(-L/\lambda_G).\qquad(\text{Eq. S12})$$

To define a lower bound to the REE efficiency, we use the width of the metallic pads on 2H-TaS2 $w_{\text{TaS}_2;\text{eff}} = 2\ \mu\text{m}$, where the flow of charge current is the strongest. Combining these results, we find

$$\gamma_{\text{REE}} \geq \frac{R_{\text{REE}}}{s}\frac{w_{\text{TaS}_2;\text{eff}}}{P\rho_{5LG}\,\lambda_G\exp(-L/\lambda_G)}.\qquad(\text{Eq. S13})$$

In calculating $s$, we used the measured bulk resistivity of 2H-TaS2 and gate-dependent 5LG resistivity $\rho_{5LG}$. Ratio $s$ changes with gate voltage and temperature, at charge neutrality it has the smallest value. In our measurement, $s$ ranges from 0.12 (50 K, $V_g = 0$ V) to 0.60 (293 K, $V_g = 55$ V). With temperature increases, $s$ increases. For a gate voltage of +35 V, we find $s =$ 0.21, 0.32, 0.40, 0.51 for temperatures of 50 K, 100 K, 150 K and 293K.

We comment on the exponential decay factor in EquationS8 that encodes the spin relaxation in the spin diffusion channel (5LG). In the actual device (see Fig. 1a, main text), the spins transverse two different regions before hitting the Co detector (FM3). The length of these regions is $y$ dependent. The heterojunction region, here approximated as a 1D strip, has a length of $L_1 = 4.6$ μm, while the 5LG channel has a length of $L_2 = 1.5$ μm. $L = L_1 + L_2 = 6.1$ μm is the total path length (largely $y$-independent). Because the spin diffusion length at the heterojunction is shorter than in the 5LG channel ($\lambda_1 < \lambda_2 = \lambda_G$), the 1D decay in Eq. S8 (and Eq. S12) using $\lambda_G$ underestimates the charge-to-spin conversion efficiency.

## 11. Relativistic first-principles calculations: additional results

We report additional results on the relativistic band structure of 2H-TaS$_2$-graphene. Figure S10 shows the DFT band structure of monolayer 2H-TaS$_2$. Monolayer 2H-TaS$_2$ is metallic as one band crosses the Fermi level. When spin-orbit coupling (SOC) is included in the calculation (panel b), new splittings occur in the band structure. For example, the metallic band that crosses the Fermi level splits by 0.29 eV at the K-point, while the splitting at the Γ point is negligible.

Figure S11 shows the DFT band structure of the 2H-TaS$_2$-graphene heterostructure (the supercell is shown in Fig. 4a, main text). Here, red dots denote bands that are predominantly located on carbon atoms and blue dots denote bands that are located on 2H-TaS$_2$ layers (dot size is proportional to magnitude of the projection). We observe that interfacial effects strongly modify the multilayer graphene band structure. Most noticeably, graphene bands located near the Fermi level exhibit a giant Rashba effect, with spin gaps reaching values as large as 20 meV (electrons) and 70 meV (holes). Despite the metallic character of 2H-TaS$_2$ (and associated strong proximity effect), the Fermi level shift with respect to the conical point of graphene is small ($\Delta E \approx -10$ meV) due to the substantial electrostatic potential energy change generated at the 2H-TaS$_2$-graphene interface, which precisely compensates the work function difference between the two systems (Refs.39,40 in main text). The Fermi level shift obtained in our DFT calculation is consistent with the prediction of the general model for graphene-metal surfaces developed in Ref. 40.

Figure S12 shows the spin texture of the two graphene bands near the Fermi level in a monolayer TaS$_2$-monolayer graphene heterostructure. In agreement with the previous calculation for the 2H-TaS$_2$-5LG heterostructure, we find that the expectation value of the in-plane spin circularly winds around the K-point of the supercell Brillouin zone and that the spins of the two bands point in opposite directions. Similar results were obtained for a bilayer TaS$_2$-bilayer graphene heterostructure.

## 12. Tight-binding model and Edelstein efficiency: additional details and analysis

The graphene multilayer with ABABA stacking was described by the Slonczewski-Weiss-McClure model of bulk graphite (Ref.58 in main text) supplemented with interface-induced Rashba interaction

$$H_{5LG} = \sum_{l=1\ldots5} \left\{ \sum_{<i,j>} c^\dagger_{l,i} \left[ -\gamma_0 + \frac{2i\lambda_l}{3} (\vec{d}_{ij} \times \vec{s}) \right] c_{l,j} + \sum_i c^\dagger_{l,i} V_{l,i} c_{l,i} \right\} + V_{\text{inter}}(\gamma_1, \gamma_3, \gamma_4) ,$$

(Eq. S18)

where $c^\dagger_{l,i} \equiv (c^\dagger_{l,i,\uparrow}, c^\dagger_{l,i,\downarrow})$ adds electrons on site $i$ in layer $l = 1,\ldots,5$, $\gamma_0$ is the intralayer nearest-neighbor hopping, and $\lambda_l$ is the Rashba SOC strength. Here, $\vec{d}_{ij}$ is the unit vector along the line segment connecting sites $ij$, and $\vec{s}$ is the vector of Pauli matrices. Interlayer terms are incorporated in $V_{\text{inter}}$ and consist of a vertical interlayer hopping ($\gamma_1$) between dimer orbitals and skewed interlayer hoppings ($\gamma_3, \gamma_4$). A layer-dependent on-site electrostatic potential (term $V_{l,i}$) is included to model the proximity-induced electric field build up across 5LG. Starting from standard values for graphite, the parameters are varied until the spectrum reproduces the DFT energy bands of states predominantly localized on carbon atoms. The parameters of the minimal model are listed in Table S1. The reference values for bulk graphite are obtained from Ref.58 in main text. To improve the agreement with the relativistic DFT band structure, the on-site energy of dimer orbitals is shifted by $\Delta'$. Except for the skewed hopping $\gamma_4$ (describing interlayer coupling between dimer and non-dimer sites), the hopping terms are in the same range for multilayers and bulk graphite, which confirms that the electronic structure of 5LG (and in particular the Dirac character of low-energy states) is overall preserved.

**Table S1** (energies in eV)

|  | **Bulk graphite** | **5LG-TaS$_2$** |
|---|---|---|
| $\gamma_0$ | 3.2 | 2.6 |
| $\gamma_1$ | 0.39 | 0.37 |
| $\gamma_3$ | 0.315 | 0.252 |
| $\gamma_4$ | 0.064 | 0.264 |
| $\Delta'$ | 0.051 | 0.018 |
| $V_{1,2,3,4,5}$ | - | $\{0.137, -0.01, -0.02, -0.06, -0.09\}$ |
| $\lambda_{1,2,3,4,5}$ | - | $\{-0.02, -0.01, 0.00, 0.00, 0.00\}$ |

The low-energy spectrum of pristine 5LG consists approximately of two bilayer-*like* quadratic bands and one monolayer-*like* linear band in both electron and hole sectors. In the 5LG-2H-TaS$_2$ heterostructure, the proximity-induced field across the graphene layers pushes away linear bands (and quadratic Dirac bands with smaller effective mass) from the Dirac point, generating large electron-hole asymmetry. Moreover, the spin degeneracy is effectively removed due to the sizeable Rashba effect, leading to spin-split states endowed with helical spin texture. Figure S13 compares the tight-binding (TB) spectrum (blue and red solid lines) against DFT energy bands shown in background (grey dots indicate states predominantly localized on carbon atoms). The green area highlights the gapping out of lighter quadratic Dirac bands due to proximity-induced electric field (note that the linear bands are gapped out more strongly, $E_K \approx 0.14$ eV for electrons and $E_K \approx -0.10$ eV for holes). Notably, holes ($E < 0$) are dramatically affected by the hybridization to TaS$_2$ orbitals. As such, the simple TB model fails to describe quantitatively the spectrum in the interval $[-0.01, -0.1]$ eV, including the appearance of giant spin splitting (with $\Delta E_s$ up to 70 meV). On the other hand, the TB model is accurate for $E > -0.01$ eV (and also below $-0.2$ eV). Trigonal warping effects and even the wavevector dependence of spin splittings are fairly well described.

The TB model can be further optimized to match the DFT spectrum of electron states up to energies $E \approx 0.9$ eV within numerical precision.

The pronounced electron-hole (e/h) asymmetry is a consequence of imperfect screening by graphene layers, which allows the penetration of the proximity-induced electric field (the same effect responsible for gap opening in bilayer/trilayer graphene adsorbed on metal surfaces.[S17, S18]

To determine the response of the heterostructure to a DC electric field, we solve the Boltzmann transport equations incorporating electron-impurity scattering within the standard relaxation time approximation. The DC charge conductivity and charge current-spin density susceptibility read as

$$\sigma_{\text{dc}}(\mu, T) = -e^2 \sum_{n=1}^{20} \iint \frac{d^2\boldsymbol{k}}{(2\pi)^2} \left(\frac{df(\varepsilon, \mu, T)}{d\varepsilon}\right)_{\varepsilon_n(\boldsymbol{k})} \tau_n(\boldsymbol{k}) \langle \hat{v}_y \rangle_{\boldsymbol{k}n} \langle \hat{v}_y \rangle_{\boldsymbol{k}n} \quad , \quad \text{(Eq. S19)}$$

$$\chi_{\text{xy}}(\mu, T) = -\frac{e\hbar}{2} \sum_{n=1}^{20} \iint \frac{d^2\boldsymbol{k}}{(2\pi)^2} \left(\frac{df(\varepsilon, \mu, T)}{d\varepsilon}\right)_{\varepsilon_n(\boldsymbol{k})} \tau_n(\boldsymbol{k}) \langle \hat{s}_x \rangle_{\boldsymbol{k}n} \langle \hat{v}_y \rangle_{\boldsymbol{k}n} \quad , \quad \text{(Eq. S20)}$$

respectively, where $f(\varepsilon, \mu, T)$ is the Fermi-Dirac distribution function, $\hat{v}_y = (i/\hbar)[H_{5\text{LG}}, \hat{y}]$ is the velocity operator along the direction of the electric field (taken along the y axis), $\hat{s}_x$ is the x-spin polarization operator and $\tau_n(\boldsymbol{k})$ is the transport relaxation time. The sum is over all states (*i.e.* 5(layers) × 2(sublattices) × 2(spins) = 20) of the 5LG tight-binding Hamiltonian. $\varepsilon_n(\boldsymbol{k})$ and the expectation values are computed *via* exact numerical diagonalization on a fine $\boldsymbol{k}$-space grid with cutoff momentum $|\boldsymbol{k}|_{\text{cut}} = 0.186$ nm$^{-1}$. The REE charge-to-spin conversion efficiency is controlled by the Edelstein ratio

$$\beta_{\text{EE}} = \sigma_{yy}/\chi_{xy} \quad . \quad \text{(Eq. S21)}$$

The latter quantity was found to be independent on the functional form of $\tau_n(\boldsymbol{k})$ to a very good approximation in monolayer graphene (Ref.[28], main text). We set a constant relaxation time in our calculations $\tau_n(\boldsymbol{k}) = \tau$ and checked that alternative functional forms led to qualitatively identical results.

Figure S14 shows the interfacial ('top') graphene layer contribution to the microscopic figure of merit calculated for the full DFT-parameterized TB model and a heuristic TB model with $V_{1,2,3,4,5} = 0$ to minimize electron-hole asymmetry and thus enforce a better agreement with experiment (see discussion above). The top layer contribution was extracted by projecting the spin operator onto the subspace of the first graphene layer. The DFT and heuristic models share the same overall behavior, *i.e.* a positive/negative Edelstein response for electrons/holes shows a tendency for saturation at high electronic density with moderate temperature dependence – all essential features reported in the main text. The heuristic model furthermore reproduces the observed monotonic increase of the Rashba-Edelstein efficiency at low electronic density as well as its saturation at increasing electronic density. The behavior of the calculated Edelstein efficiency is therefore consistent with the charge-spin conversion characteristics in the experiment, providing yet further evidence in favor of spin-helical 2D Dirac fermions. Importantly, these findings are robust against reasonable variations in the TB parameters, *e.g.* the microscopic figure of merit remains virtually unchanged by modifying the distribution of Rashba SOC within the first two layers to $\lambda_1 = 0.03$ eV and to $\lambda_2 = 0.00$ eV. It is important to note that $\beta_{EE}$ at saturation is $\approx 20\%$, higher for electrons than for holes, whilst the measured efficiency (at room temperature) shows much larger difference ($|\gamma_{REE}| \approx 1.4\%$ for holes *versus* $|\gamma_{REE}| \approx 4.3\%$ for electrons). The microscopic figure of merit $\beta_{EE}$ and the charge-to-spin conversion efficiency of the device ($\gamma_{REE}$) are connected by $\gamma_{REE} = \vartheta_{2D} \times \beta_{EE}$, where $\vartheta_G = D_G \nu_G / \lambda_G \nu_{2D}$ is the spin-channel efficiency. $\nu_G(\nu_{2D})$ and $D_G$ are, respectively, the 2D density of states and spin diffusion constants of 5LG (5LG/2H-TaS$_2$). Figure S15 shows the experimental data for $D_G/\lambda_G$ at selected temperatures. The gate-voltage dependence is relatively weak, with $\vartheta_{2D}$ attaining the largest values for electrons.

## Supplementary References

**S1.** Yang, Y. *et al*, P. Enhanced Superconductivity upon Weakening of Charge Density Wave Transport in 2H-TaS2 in the Two-Dimensional Limit. *Phys. Rev. B* **2018**, *98*, 035203.

**S2.** Maasen, T.; Vera-Marun, I. J.; Guimaraes, M. H. D.; van Wees, B. J. Contact-Induced Spin Relaxation in Hanle Spin Precession Measurements. *Phys. Rev. B* **2012**, *86*, 235408.

**S3.** Han, W. *et al*. Tunneling Spin Injection into Single Layer Graphene. *Phys. Rev. Lett.* **2010**, *105*, 167202.

**S4.** Idzuchi, H.; Fukuma, Y.; Takahashi, S.; Maekawa, S.; Otani, Y. Effect of Anisotropic Spin Absorption on the Hanle Effect in Lateral Spin Valves. *Phys. Rev. B* **2014**, *89*, 081308.

**S5**. No, Y. *et al*. Layer Number Identification of CVD-Grown Multilayer Graphene Using Si Peak Analysis. *Sci. Rep*. **2018**, *8*, 571.

**S6.** Nguyen, T. A.; Lee, J. -U.; Yoon, D.; Cheong, H. Excitation Energy Dependent Raman Signatures of ABA- and ABC-Stacked Few-Layer Graphene. *Sci. Rep*. **2014**, *4*, 4630.

**S7.** Sasaki, T.; *et al*. Evidence of Electrical Spin Injection into Silicon Using MgO Tunnel Barrier. *IEEE Trans. Magn*. **2010**, *46*, 1436-39.

**S8.** Donahue, M. J.; Porter, D. G. *Interagency Report NISTIR 6376*, National Institute of Standards and Technology, Gaithersburg, MD **1999**.

**S9.** Hamrle, J. *et al*. Y. Determination of Exchange Constants of Heusler Compounds by Brillouin Light Scattering Spectroscopy: Application to $Co_2MnSi$. *J. Phys. D: Appl. Phys.* **2009**, *42*, 084005.

**S10.** Chikazumi; S., English edition prepared with the assistance of Graham Jr., C. D. *Physics of Ferromagnetism*. Clarendon Press, Oxford **1997**.

**S11.** Huang, C.; Chong, Y. D.; Cazalilla, M. A. Anomalous Nonlocal Resistance and Spin-Charge Conversion Mechanisms in Two-Dimensional Metals. *Phys. Rev. Lett.* **2017**, *119*, 136804.

**S12.** Bludov, Y. V.; Ferreira, A.; Peres, N. M. R.; Vasilevskiy, M. I. A Primer on Surface Plasmon-Polaritons in Graphene. *Int. J. Mod. Phys. B* **2013**, *27*, 1341001.

**S13.** Nozieres, P. Cyclotron Resonance in Graphite, *Phys. Rev.* **1958**, *109*, 1510-1521.

**S14.** I. Crassee, *et al*. Multicomponent Magneto-Optical Conductivity of Multilayer Graphene on SiC. *Phys. Rev. B* **2011**, *84*, 035103.

**S15.** Crassee, I. *et al*, A. B. Giant Faraday Rotation in Single- and Multilayer Graphene. *Nat. Phys.* **2011**, *7*, 48-51.

**S16.** Hong, X.; Zou, K.; Zhu, J. Quantum Scattering Time and Its Implications on Scattering Sources in Graphene. *Phys. Rev. B* **2009**, *80*, 241415(R).

**S17.** Zheng, J. *et al* Interfacial Properties of Bilayer and Trilayer Graphene on Metal Substrates. *Sci. Rep.* **2013**, *3*, 2081.

**S18.** Khan, E., Rahman, T. S.; Subrinaa, S. Electronic Structure of Bilayer Graphene Physisorbed on Metal Substrates. *Journal of Applied Physics* **2006**, *120*, 185101.

**Supplementary Figures**

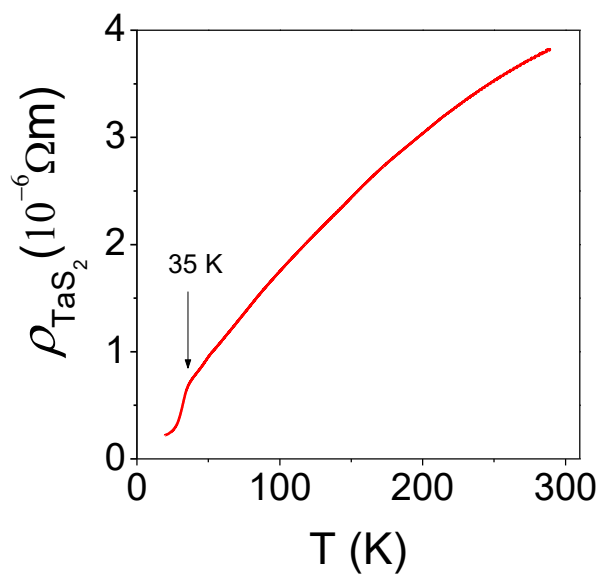

**Figure S1.** Characterization of exfoliated 12 nm thick thin film 2H-TaS$_2$. Temperature dependent resistivity of 2H-TaS$_2$, decreases with temperature. At 35 K the resistivity changes abruptly (black arrow), indicating the formation of a charge density wave.

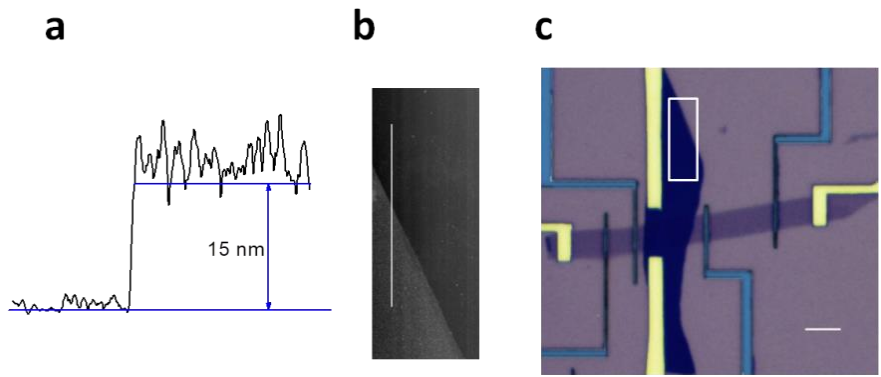

**Figure S2.** Atomic force microscopy (AFM) measurement of TaS$_2$ thickness. (a) Thickness of the 2H-TaS$_2$ layer measured by AFM. (b) AFM image taken from the area indicated by the box in (c). The line shows the line cut direction for the height (thickness) measurement shown in (a). (c) False-colored optical image of the device. Scale bar: 5 µm.

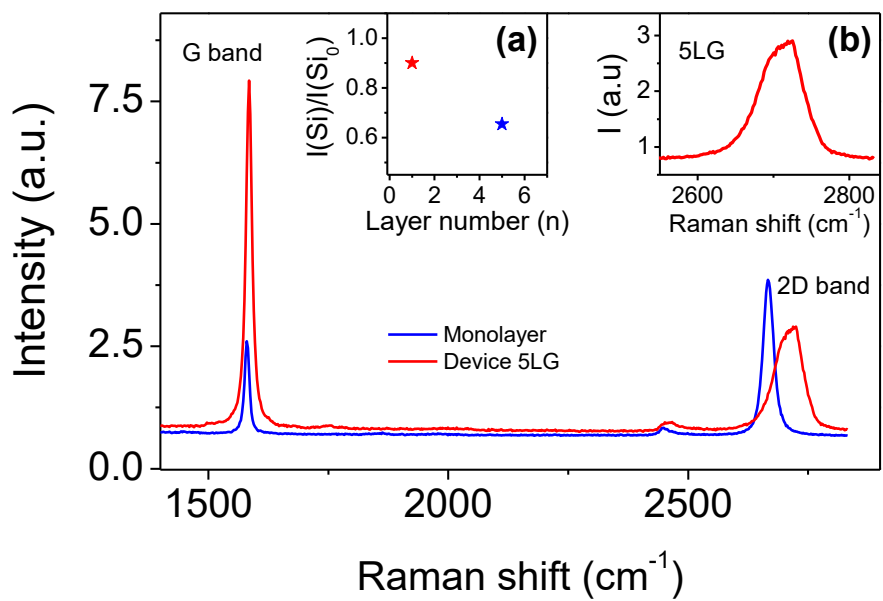

**Figure S3.** Raman spectra of 5LG in our device. Inset (a): Si Raman peak intensity ratio for 5LG used in device and a monolayer graphene. Inset (b): Zoom in of the 2D peak of 5LG. The relative intensity of the Raman peak shows that the number of graphene layer is five while the shape of the 2D peak indicates that the layer sequence is ABABA.

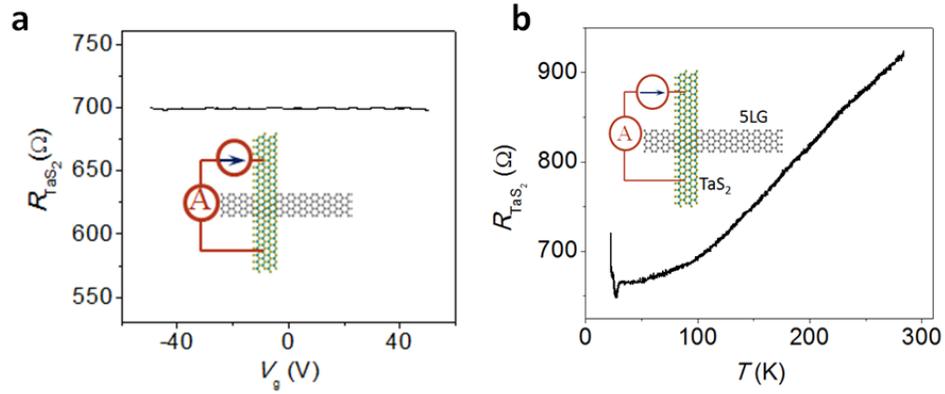

**Figure S4.** Resistance of the TaS$_2$ thin film including the heterostructure region as a function of gate voltage and temperature. (a) TaS$_2$ two probe resistance $R_{TaS_2}$ measured by applying a bias voltage and register the current upon sweeping the gate voltage. (b) From 50 K to 300 K, $R_{TaS2}$ shows metallic behavior. Below 27 K, $R_{TaS_2}$ increases sharply, indicating the formation of charge density wave. Measurements were performed by two-probe method shown in the inset.

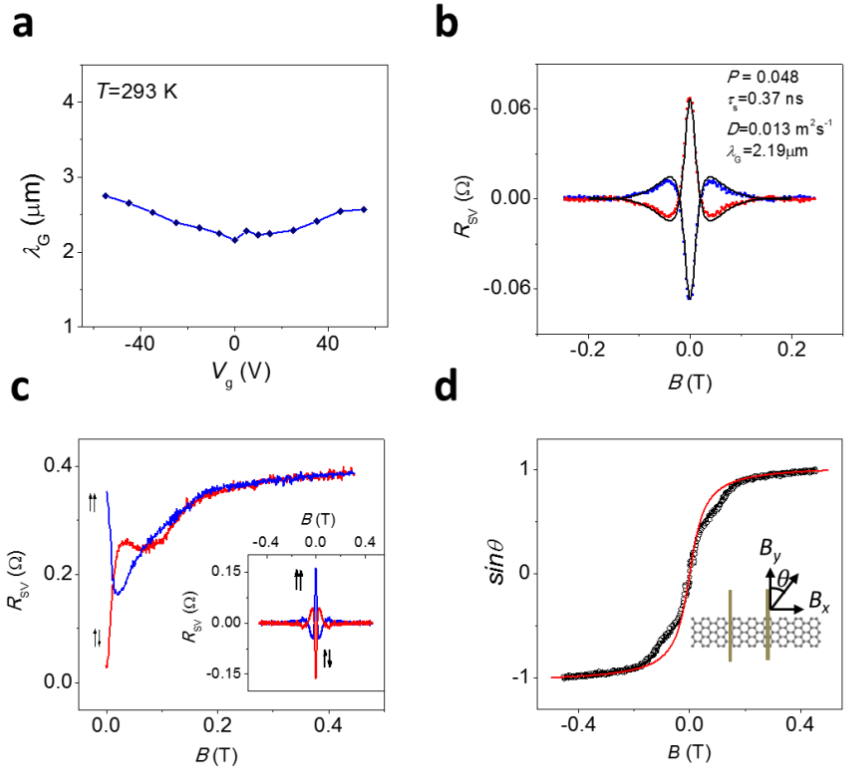

**Figure S5.** Spin transport in 5LG and magnetization of Co electrode. (a) Gate voltage dependence of spin diffusion length obtained from out of plane Hanle measurement at 293 K. (b) In-plane Hanle measurement at $T = 293$ K. The spin diffusion length extracted from the in-plane Hanle curve is comparable to that from the out-of-plane Hanle curve (main Figure 1f). (c) In-plane Hanle curve measured at 22 K. The inset shows the symmetrized curves. (d) Co magnetization rotation obtained from the results in (c). The red curve is a fit to $\tanh(B/B_{sat})$, $B_{sat} \approx 0.2$ T.

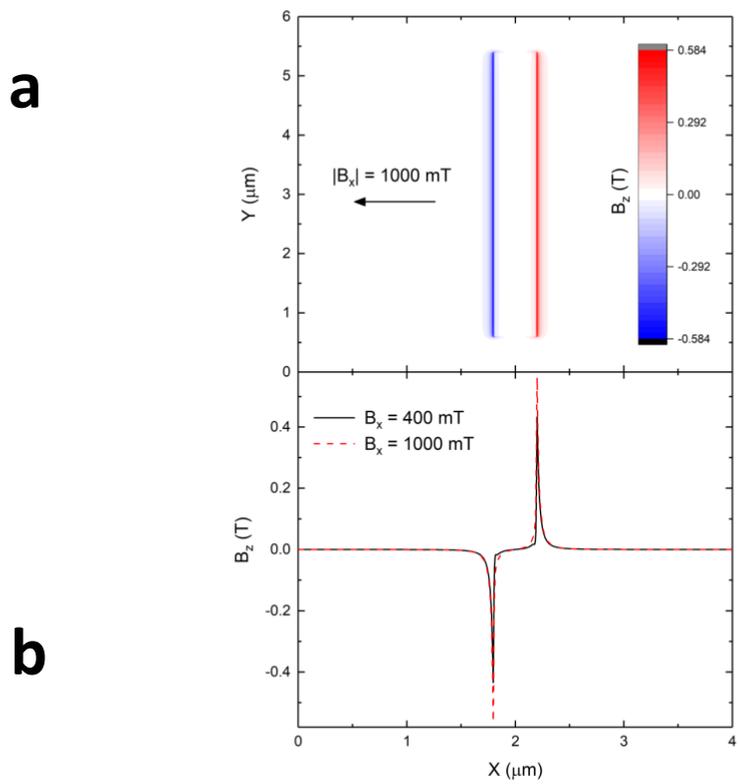

**Figure S6.** Stray field map of Co (FM3) electrode. (a) Demagnetization (stray) field map ($B_z$) in the *x-y* plane at *z* = 0 nm for an applied field $B_x$ = -1000 mT. (b) Profile of $B_z(x)$ for two values of the applied field as shown.

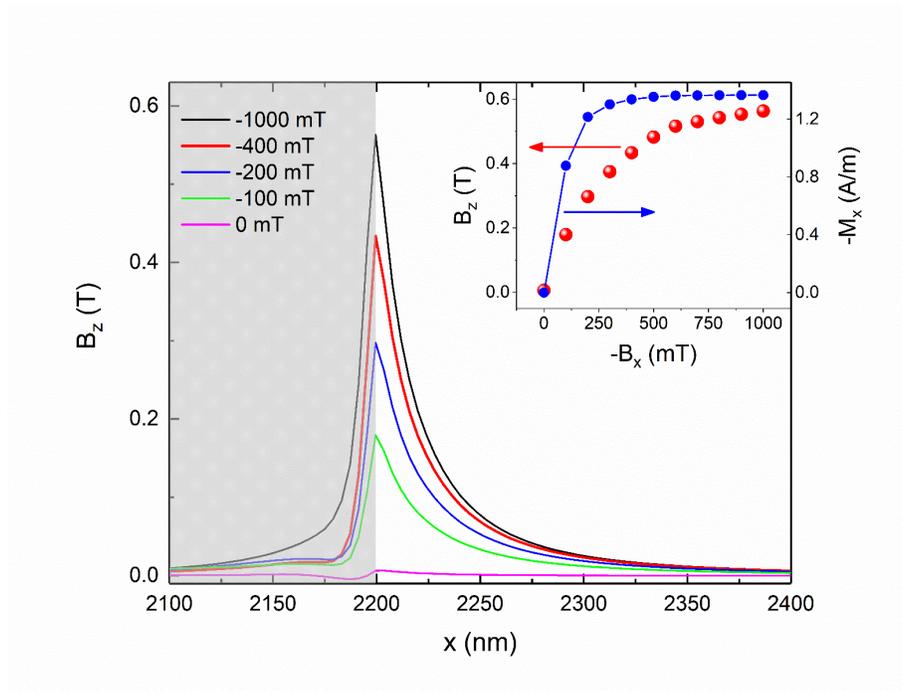

**Figure S7.** z-component of stray field. Demagnetizing (stray) field profile $B_z$ at the right-hand edge of the structure for several values of the applied field. The grey area indicates the extent in $x$ of the magnetic structure with $B_z$ decaying away from it. Inset: Peak in $B_z(x)$ and $M_x(x)$ as a function of the applied field.

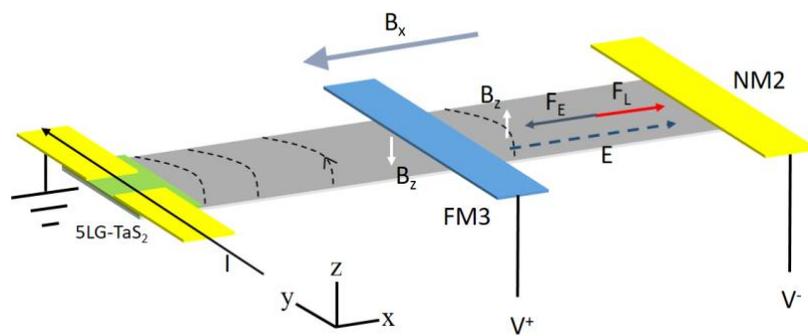

**Figure S8.** OHE induced by stray field. Schematic of Ohmic field lines emanating from current injector and associated Hall voltage drop between FM3 and NM2.

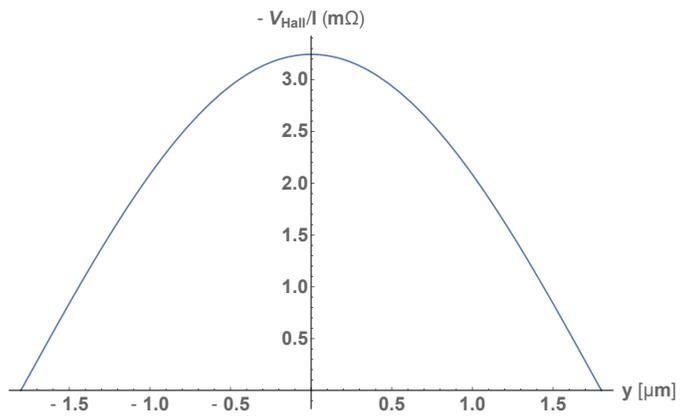

**Figure S9.** Ordinary Hall effect due to stray field. Ordinary Hall resistance (= $V_H$ / I) between contacts FM3 and NM2 as a function of *y* coordinate.

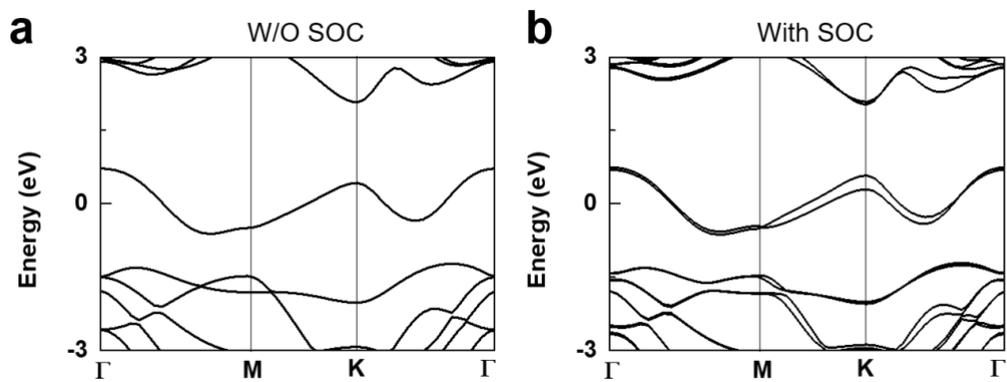

**Figure S10.** DFT band structure of 2H-TaS$_2$ monolayer without (a) and with spin-orbit couplings (b). The Fermi levels are set to zero.

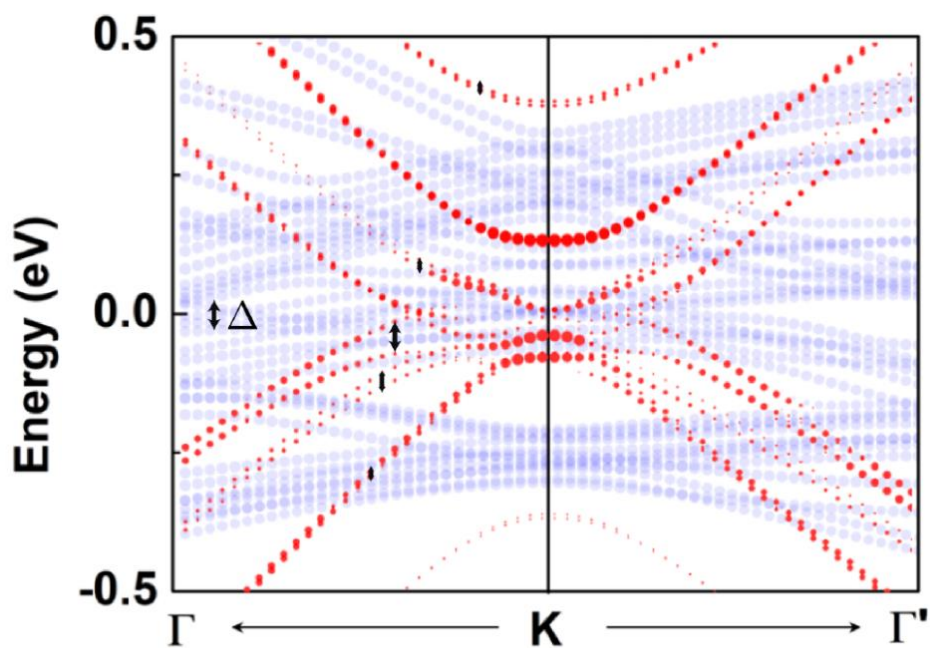

**Figure S11.** Projected DFT band structure of a multilayer graphene-TaS$_2$ heterostructure including spin-orbit interactions. Red (blue) dots denote bands with wavefunctions that are predominantly localized on graphene (TaS$_2$) layers. For this, the wavefunctions of the multilayer were projected onto atomic orbitals. The size of the red (blue) dots is proportional to the sum of the absolute magnitudes of the projections onto carbon (Ta and S) atomic orbitals. Representative spin gaps are indicated with double arrows. The Fermi level is set to zero.

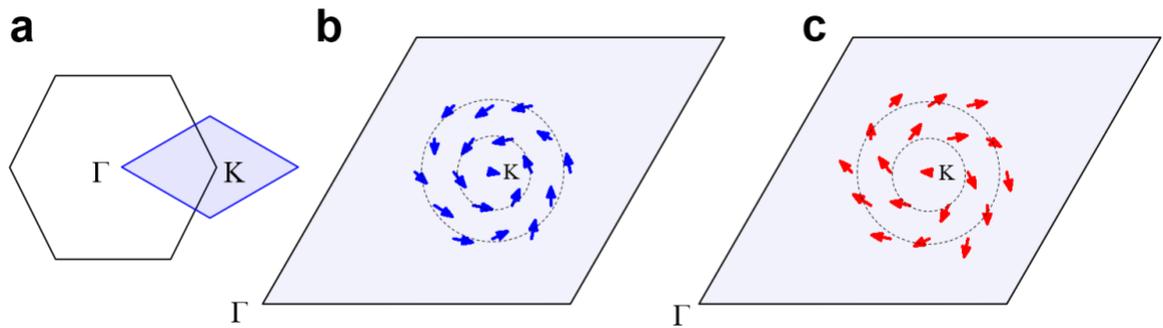

**Figure S12.** DFT spin texture of a monolayer TaS$_2$-monlayer graphene heterostructure. (a) The Brillouin zone of the TaS$_2$-graphene heterostructure. (b), (c) Spin textures of the low-energy graphene bands. The Brillouin zone areas shown in (b) and (c) correspond to the blue area in (a).

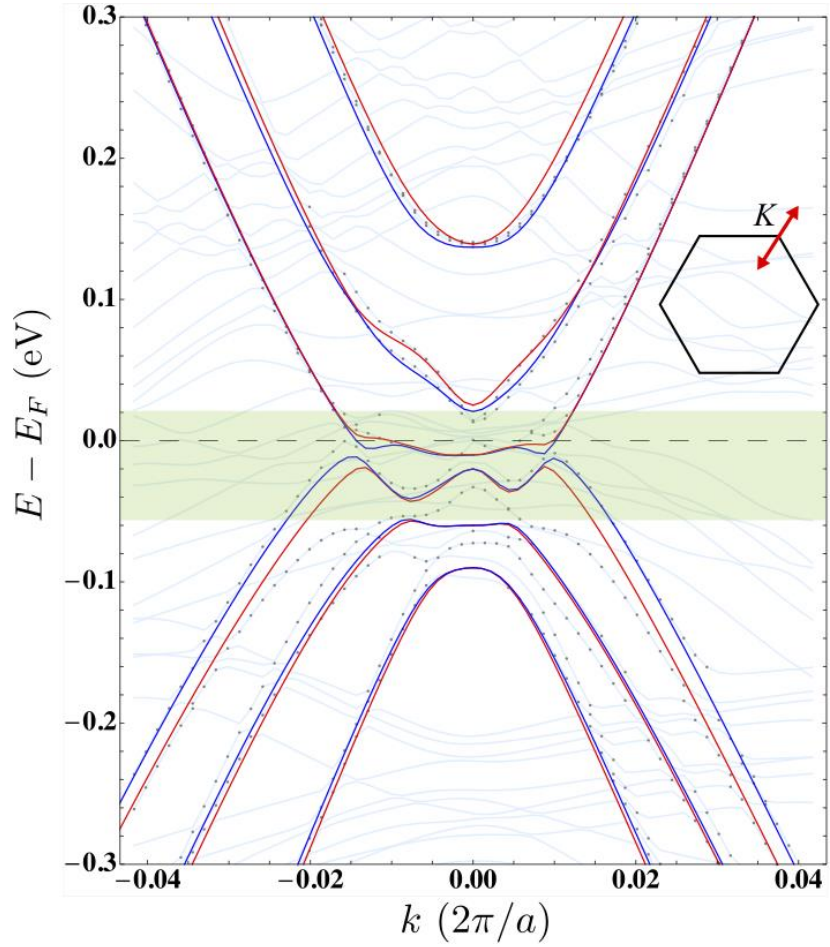

**Figure S13.** Tight-binding and DFT band structures of 5LG-TaS$_2$. Red and blue solid lines are tight-binding bands and light blue solid lines represent the full relativistic DFT results (grey dots mark states predominantly localized on graphene). The green area highlights the gapping out of lighter quadratic Dirac bands.

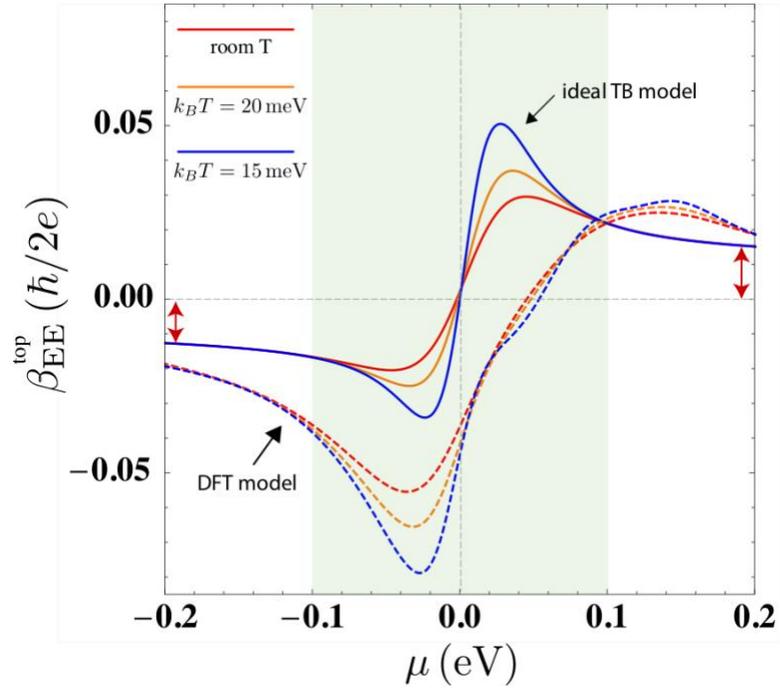

**Figure S14.** Edelstein efficiency of interfacial 'top' graphene layer: ideal model *versus* full model. Fermi energy dependence for selected temperatures computed from full (dashed) and 'ideal' (solid) tight-binding models. The regime of experiment is highlighted by the green area.

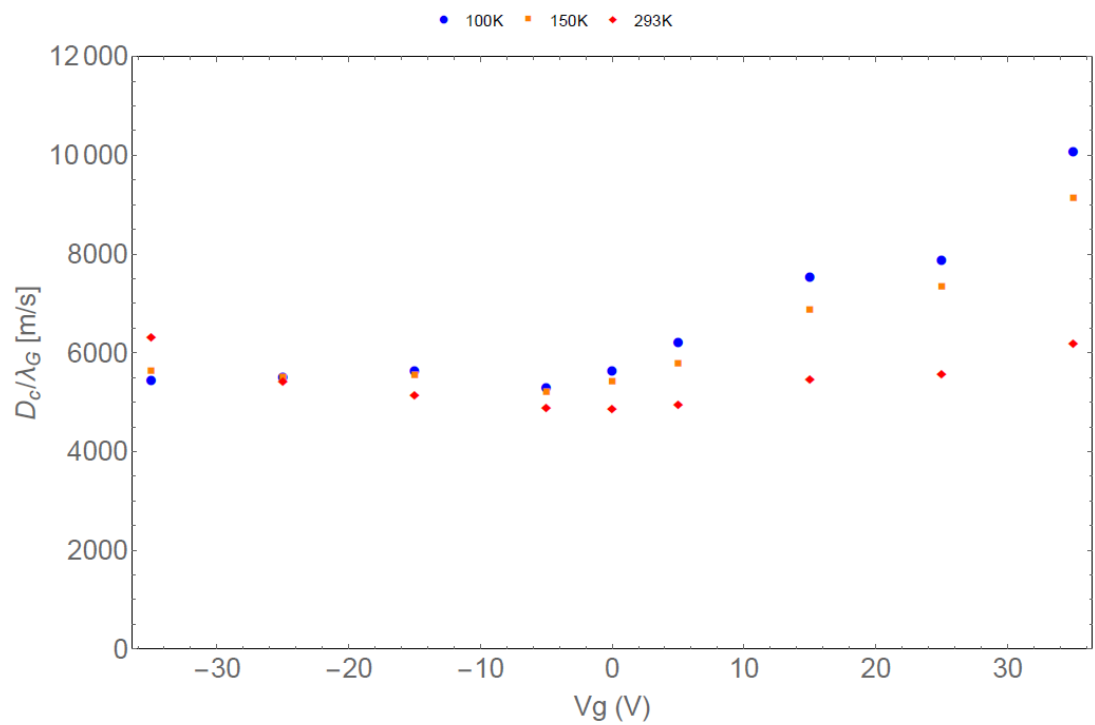

**Figure S15.** Spin-channel efficiency. Gate-voltage dependence of $D_G/\lambda_G$ at selected temperatures.